\documentclass[%
 reprint,
nofootinbib,
 amsmath,amssymb,
 aps,
floatfix,
twoside]{revtex4-1}

\usepackage[utf8]{inputenc}
\usepackage[brazilian]{babel}
\usepackage[T1]{fontenc}

\usepackage{mathtools}
\usepackage{siunitx}
\usepackage{amsmath}
\usepackage{amsfonts}
\usepackage{amssymb}
\usepackage{graphicx}
\usepackage{lipsum}
\usepackage{color}
\usepackage[colorlinks=true,citecolor=blue,linkcolor=blue,urlcolor=blue]{hyperref}
\usepackage{times}
\usepackage{amstext}
\usepackage{latexsym}
\usepackage[running,mathlines]{lineno}
\usepackage{verbatim}
\usepackage{physics}
\usepackage{bm}
\usepackage{ulem}
\usepackage[version=3]{mhchem}
\usepackage[dvipsnames]{xcolor}

\newcommand*\Diff[1]{\mathop{}\!\mathrm{d^#1}}

\begin{document}
\raggedbottom

\title{\Large{Íons Aprisionados como Arquitetura para Computação Quântica} \\
(Trapped Ions as an Architecture for Quantum Computing)}

\author{Gabriel P. L. M. Fernandes} \email[Endereço de correspondência:]{ gabrielpedro@df.ufscar.br}

\affiliation{Universidade Federal de São Carlos, Departamento de Física, São Carlos, SP, Brasil}

\author{Alexandre C. Ricardo}
\affiliation{Universidade Federal de São Carlos, Departamento de Física, São Carlos, SP, Brasil}

\author{Fernando R. Cardoso}
\affiliation{Universidade Federal de São Carlos, Departamento de Física, São Carlos, SP, Brasil}

\author{Celso J. Villas-Boas}
\affiliation{Universidade Federal de São Carlos, Departamento de Física, São Carlos, SP, Brasil}

\begin{abstract} 

Neste artigo descrevemos uma das plataformas mais promissoras para a construção de um computador quântico universal, que consiste em uma cadeia de $N$ íons aprisionados em um potencial harmônico, cujos estados internos funcionam como qubits e são acoplados a modos vibracionais coletivos da cadeia. A partir de tal acoplamento, é possível construir interações entre diferentes íons, isto é, interações qubit-qubit que, juntamente com operações individuais sobre os íons, permitem construir um computador quântico como primeiramente proposto por Cirac e Zoller  na década de 90 [Phys. Rev. Lett. \textbf{74}, 4091 (1995)]. Aqui discutimos desde a física envolvida no aprisionamento de íons em armadilhas eletromagnéticas até a engenharia de hamiltonianos necessária para gerar um conjunto universal de portas lógicas, fundamental para a execução de algoritmos quânticos mais complexos. Por último, apresentamos o atual estado da arte da computação quântica em sistemas de íons aprisionados, destacando os recentes avanços obtidos por empresas e projetos governamentais que utilizam tal arquitetura, como IonQ e AQTION.

\begin{description}
\item[Palavras-chave] Computação quântica; Íons aprisionados; Algoritmos quânticos.
\end{description}

\bigbreak 

In this paper we describe one of the most promising platforms for the construction of a universal quantum computer, which consists of a chain of $N$ ions trapped in a harmonic potential, whose internal states work out as qubits, and are coupled to collective vibrational modes of the chain. From such coupling, it is possible to build interactions between different ions of the chain, that is, qubit-qubit interactions that, together with individual operations on the ions, allow building a quantum computer as first proposed by Cirac and Zoller in the 1990s [Phys. Rev. Lett. \textbf{74}, 4091 (1995)]. Here we discuss from the physics involved in trapping ions in electromagnetic potentials to the Hamiltonian engineering needed to generate a universal set of logic gates, fundamental for the execution of more complex quantum algorithms. Finally, we present the current state of the art of quantum computing in trapped ion systems, highlighting recent advances made by companies and government projects that use such architecture, such as IonQ and AQTION.

\begin{description}
\item[Keywords] Quantum computation; Trapped ions; Quantum algorithms.
\end{description}
\end{abstract}

\maketitle

\section{Introdução}

O princípio da superposição encontra-se na origem de uma série de atributos e propriedades de estados quânticos, essenciais na descrição de sistemas físicos. Entre suas mais notáveis consequências, decorrem os fenômenos de coerência \cite{Streltsov2017} e emaranhamento \cite{Horodecki2009}, ambos fundamentais para o entendimento da natureza de propriedades não-clássicas e suas aplicações para o desenvolvimento de novas tecnologias, empregadas, por exemplo, à computação e ao processamento de informação \cite{nielsen_chuang_2010}. Tem sido demonstrado que, a partir desses fenômenos, máquinas que operam sob as leis da Mecânica Quântica podem executar cálculos de forma mais eficiente do que aquelas que operam sob as leis da Mecânica Clássica \cite{hidary2019quantum}, o que possibilita um ganho substancial de poder computacional na realização de determinadas tarefas \cite{Montanaro2016}. A extensão do poder computacional dessas novas máquinas, denominadas computadores quânticos, depende da tarefa que se quer realizar e pode se traduzir em ganhos exponenciais no tempo de processamento \cite{deutsch1992}
ou até mesmo na resolução de problemas que não poderiam ser resolvidos em tempo hábil pelos  mais potentes supercomputadores da atualidade \cite{Arute2019}.

Em geral, computadores quânticos são construídos de forma que sejam programáveis em um entre os diferentes possíveis modelos de computação quântica \cite{Nimbe}. Em específico, considerando o modelo de portas lógicas em variáveis discretas, um computador quântico é uma rede de sistemas
cujas dinâmicas são restritas a dois níveis.  Individualmente, cada sistema é denominado \textit{qubit}, a unidade básica de memória da computação quântica na qual a informação é alocada e manipulada. 

Os \textit{qubits} diferem dos bits clássicos no seguinte aspecto: enquanto um bit apresenta comportamento estritamente determinístico, dado por um estado bem definido, $0$ ou $1$, o comportamento de um \textit{qubit} tem caráter probabilístico, podendo se apresentar nos estados $\ket{0}$ ou $\ket{1}$, que formam a base computacional da computação quântica, ou em superposições normalizadas destes \cite{Steane1998}. No modelo de computação considerado, a manipulação da informação alocada nos \textit{qubits} é realizada via aplicação de portas lógicas quânticas, que são interações aplicadas com o objetivo de rotacionar ou emaranhar os estados dos \textit{qubits} \cite{Barenco1995}. A engenharia por trás dessas interações e o mapeamento dos autoestados na base computacional são questões inevitavelmente ligadas à natureza dos sistemas que constituem o computador quântico \cite{Divincenzo2000}.

Ao longo das últimas décadas, diferentes sistemas físicos foram considerados para a implementação de computação quântica \cite{Divincenzo2000,Ladd2010}. Entre os que se mostraram mais apropriados, isto é, aqueles em que se obteve maior grau de controle da dinâmica do sistema, estão os sistemas baseados em fotônica \cite{Slussarenko2019}, em circuitos supercondutores \cite{Kjaergaard2020} e em íons aprisionados \cite{Steane1997, Haffner2008, Bruzewicz2019}, cada qual com suas particularidades. Em íons aprisionados, a primeira proposta de um computador quântico escalável foi feita por Cirac e Zoller \cite{PhysRevLett.74.4091}, que idealizaram um sistema composto por uma cadeia linear de $N$ íons confinados em uma armadilha de potencial aproximadamente harmônico, na qual os estados dos qubits são manipulados via aplicação de pulsos de laser, que acoplam os estados internos dos íons aos modos coletivos de movimento da cadeia \cite{James1998,Blatt2008}. Atualmente, com o investimento financeiro de empresas como a IonQ \cite{IonQ} e de projetos governamentais como o AQTION \cite{AQTION}, arquiteturas inspiradas na proposta inicial feita por Cirac e Zoller têm apresentado crescente sucesso comercial e novos avanços tecnológicos. Em vista disso, o intuito desse artigo é expor de forma didática a implementação de computação quântica em sistemas de íons aprisionados, discutindo alguns dos mais fundamentais resultados alcançados até o momento.

Este artigo está organizado da seguinte forma:  Na Seção \ref{sec2}, discutimos o armadilhamento de um único íon em uma armadilha de Paul, onde tratamos as equações clássicas de movimento a fim de derivar o hamiltoniano do movimento de um único íon em uma armadilha. Na Seção \ref{sec1}, consideramos dois diferentes esquemas quanto a escolha do par de estados dentre os níveis internos dos íons que serão mapeados nos estados $\ket{0}$ e $\ket{1}$, adotando o modelo matematicamente mais simples para as seções seguintes. Na Seção \ref{sec3}, consideramos um íon aprisionado interagindo com um pulso de laser ressonante ou quase-ressonante a apenas uma transição atômica tal que a dinâmica do íon esteja de fato restrita a um sistema de dois níveis. Discutimos também a engenharia necessária para construir diferentes interações que permitem controlar a dinâmica do sistema. Na Seção \ref{sec4}, consideramos o sistema escalável (e portanto mais interessante para computação quântica) de uma cadeia linear de $N$ íons submetidos a um potencial harmônico, onde os íons da cadeia interagem entre si via repulsão coloumbiana. Na Seção \ref{sec5}, construímos um conjunto de operações elementares que podem ser aplicadas nos íons da cadeia tal que quaisquer outras operações unitárias possam ser decompostas em termos destas, o que é essencial para a implementação de algoritmos quânticos. Na Seção \ref{sec6}, demonstramos como as interações previamente construídas podem ser utilizadas para a implementação do algoritmo de teletransporte quântico. Na Seção \ref{sec7}, discutimos o atual estado da arte dos computadores quânticos baseados em íons aprisionados. E, por fim, na Seção \ref{sec8}, apresentamos nossas conclusões.
 
 \section{Armadilhas de Íons}\label{sec2}
 
Armadilhas capazes de confinar íons
em uma região do espaço são essenciais para a implementação de computação quântica em plataformas de íons aprisionados. 
Nesse sentido, as formas mais conhecidas de armadilhar íons são as armadilhas de Penning e as armadilhas de Paul, resultados dos trabalhos independentes dos físicos Hans Dehmelt \cite{Dehmelt1968,Dehmelt1969} e Wolfgang Paul \cite{Paul1990}, que dividiram o prêmio Nobel de Física de 1989 pelo desenvolvimento das técnicas de armadilhamento utilizadas até os dias atuais \cite{nobel}.  Armadilhas de Penning utilizam campos eletrostáticos e magnéticos para tal fim enquanto armadilhas de Paul utilizam apenas campos elétricos - estáticos e oscilantes. Dada sua crescente popularidade em trabalhos de computação quântica, o confinamento de íons em uma armadilha de Paul será o objeto de estudo dessa seção. Posteriormente, trataremos o problema do confinamento de uma cadeia linear de $N$ íons em uma armadilha, mas, por ora, dedicaremos nossos esforços ao armadilhamento de um único íon. Para isso, consideraremos o potencial elétrico parcialmente dependente do tempo \cite{Blatt2003}
\begin{equation}
\label{eq1}
      \Phi(\vec{x},t)=\frac{U}{2}\sum_{i=1}^{3}\alpha_i x_i^2 + \frac{\tilde{U}}{2}\cos(\omega_{\textrm{rf}}t)\sum_{i=1}^{3}\tilde{\alpha}
     _{i} x_i^2,
\end{equation}
\noindent onde $(x_1,x_2,x_3) = (x,y,z)$ denotam as coordenadas cartesianas, $U$ $\big(\tilde{U}\big)$ denota a intensidade do potencial independente (dependente) do tempo com dimensão apropriada e as constantes $\alpha_i$ e $\tilde{\alpha}_i$ são parâmetros geométricos da armadilha. Estamos supondo que a frequência do campo oscilante, $\omega_{\textrm{rf}}$, está na faixa das radiofrequências.
  
  Como discutido nos cursos básicos de eletromagnetismo \cite{griffiths-em}, o potencial elétrico deve satisfazer à equação de Laplace, 
  \begin{equation}
      \nabla^2\Phi(\vec{x},t)=0,
  \end{equation}
 \noindent o que implica que o confinamento do íon na armadilha é dinâmico, uma vez que não existe ponto de mínimo global no potencial, apenas pontos de sela. Além disso, a equação de Laplace gera restrições aos valores dos parâmetros geométricos, isto é,
 \begin{equation}
 \begin{split}
 \begin{cases}
     \alpha_1 + \alpha_2 + \alpha_3& =  0, \\ 
\tilde{\alpha}_1 + \tilde{\alpha}_2 + \tilde{\alpha}_3& =  0,
 \end{cases}
\end{split} 
 \end{equation}
\noindent sendo possível, portanto, construir armadilhas com diferentes configurações geométricas, dependendo da escolha dos parâmetros $\alpha_i$ e $\tilde{\alpha}_i$.

Como o potencial é separável em coordenadas cartesianas, o tratamento das equações clássicas de movimento se resume a determinar a trajetória do íon sob a ação da força gerada pelo potencial elétrico em uma direção e generalizar o resultado para as demais. Assim, para a direção axial, temos a equação diferencial ordinária
\begin{equation} \label{d2x}
     \dv[2]{x}{t}= -\frac{Z|e|}{m}\left[U\alpha_x + \tilde{U}\tilde{\alpha}_x\cos(\omega_{\textrm{rf}})\right]x,
\end{equation}
 \noindent onde $Z$ é o grau de ionização do átomo (número de prótons do íon), $m$ é a massa do íon e $e$ é a carga eletrônica. Introduzindo os parâmetros
 \begin{align}
\xi&=\frac{\omega_{\textrm{rf}}t}{2}, &  a_x &= \frac{4Z|e|U\alpha_x}{m\omega_{\textrm{rf}}^2},  &  q_{x}&= - \frac{2Z |e| \tilde{U}\tilde{\alpha}_x}{m\omega_{\textrm{rf}}^2},
\end{align}
 \noindent a equação (\ref{d2x}) pode ser escrita como
 \begin{equation}
     \dv[2]{x}{\xi}+\left[a_x - 2q_x\cos(2\xi)\right]x=0,
 \end{equation}
 \noindent que é conhecida como equação diferencial de Mathieu e possui soluções conhecidas \cite{mathieu}. No regime em que $a_x,q_x^2 \ll 1$ \cite{Blatt2003}, é possível demonstrar que 
 \begin{equation} \label{axq2x}
     x(t)\approx A \cos(\nu_x t)\left[1-\frac{q_x}{2}\cos(\omega_{\textrm{rf}}t)\right],
 \end{equation}
  \noindent onde $A$ é a amplitude de oscilação do íon na armadilha e
 \begin{equation}
     \nu_x = \sqrt{a_x+\frac{q_x^2}{2}}\frac{\omega_{\textrm{rf}}}{2}.
 \end{equation}

\begin{figure}[t]
\includegraphics[width=1.0\linewidth]{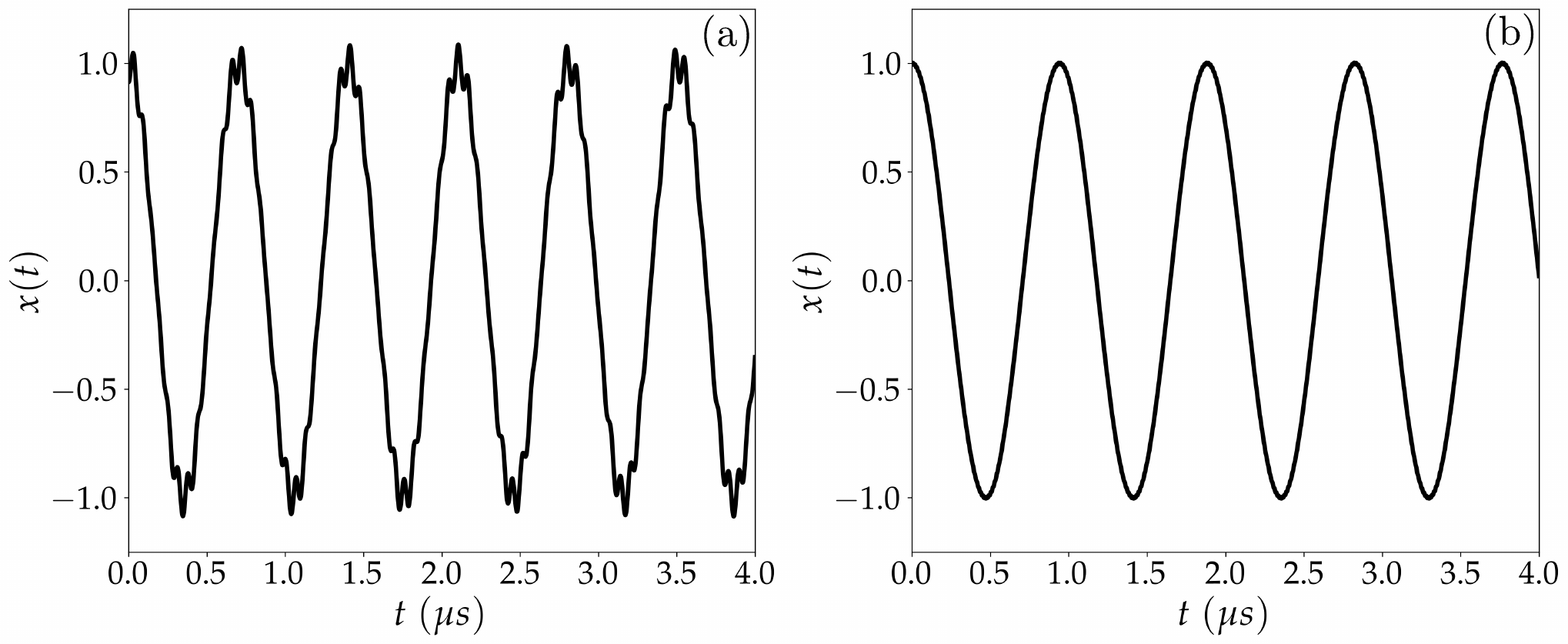}
\caption{\label{x_t} Gráfico da posição do íon em função do tempo para valores de (a) $\omega_{\textrm{rf}}=0,1 \ \textrm{GHz}$ e (b) $\omega_{\textrm{rf}}=0,5 \ \textrm{GHz}$, no regime da equação (\ref{axq2x}), com $q_x = 0.17$, $\nu_x = 8.9\textrm{ MHz}$. Pode-se perceber que, para valores de $\omega_{\textrm{rf}}$ suficientemente grandes, como ocorre em (b), o movimento do íon se aproxima de um movimento harmônico, sendo então possível desprezar o termo de micromovimento, representado pela visível deformação em relação ao movimento harmônico nos picos e vales de (a), no limite de altas frequências.}
\end{figure}

As expressões acima descrevem o movimento do íon na direção axial: oscilações harmônicas de frequência $\nu_x$ (movimento secular ou macromovimento) sobrepostas por oscilações de menor amplitude e maior frequência (micromovimento). Como mostra a Figura \ref{x_t}, para frequências do potencial oscilatório suficientemente altas, na faixa das radiofrequências, o micromovimento pode ser desprezado, aproximando a trajetória do íon na armadilha para um movimento harmônico na direção axial, o que justifica a decisão de denotar a frequência do potencial oscilatório como $\omega_{\textrm{rf}}$ desde o início.

Desta forma, no limite de altas frequências, a hamiltoniana do movimento do íon em uma dimensão é dada por
\begin{equation}\label{hmclass}
    H_{m} = \frac{p^2}{2m} + \frac{m}{2}\nu_{x}^{2}x^2,
\end{equation}
\noindent isto é, a hamiltoniana do oscilador harmônico livre. 

Quanticamente, ainda no limite de altas frequências, é possível aproximar o hamiltoniano do movimento do íon em uma dimensão para o hamiltoniano de um oscilador harmônico de frequência $\nu_x$,
trocando a coordenada $x$ e o momento $p$ pelos operadores $\hat{x}$ e $\hat{p}$, que satisfazem a relação de comutação $\comm{\hat{x}}{\hat{p}}=i\hslash$. Para isso, a dependência temporal do potencial elétrico pode ser considerada uma perturbação cujos efeitos, no regime em consideração, não alterarão significativamente os autoestados do hamiltoniano do oscilador harmônico.

 \section{Escolha dos Níveis Internos dos Íons}\label{sec1}

Em arquiteturas de íons aprisionados, as interações que permitem manipular a informação alocada nos qubits são construídas de acordo com a teoria da interação radiação-matéria \cite{scully, walls2007quantum}, aplicando-se pulsos de laser com frequências ressonantes ou quase-ressonantes a apenas uma transição atômica do íon, tal que todos os outros níveis possam ser desprezados, permitindo assim que os dois estados internos da transição selecionada sejam mapeados na base computacional, isto é, reconhecidos como $\ket{0}$ e $\ket{1}$. A escolha de quais níveis internos devem ser mapeados na base computacional é feita considerando a paridade das funções de onda desses estados, que influenciam na engenharia por trás das interações que são aplicadas ao sistema \cite{lloyd}, conforme será explicitado nas próximas seções.

\begin{figure}[h]
    \centering
    \includegraphics[scale = 1.15]{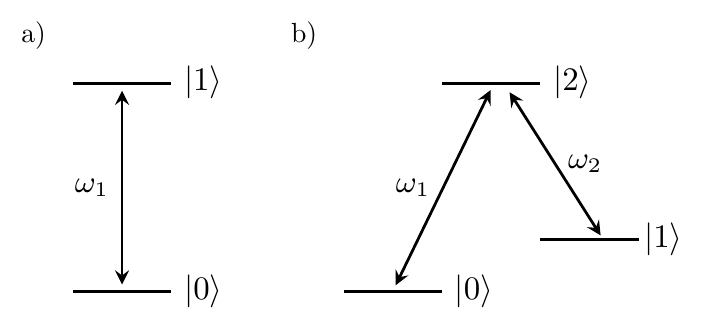}
    \caption{Diagrama dos níveis de energia segundo o esquema de (a) dois níveis e (b) três níveis. Em (a), os estados $\ket{0}$ e $\ket{1}$ têm paridades distintas. Em (b), os estados $\ket{0}$ e $\ket{1}$ têm paridades iguais e é portanto necessário introduzir um estado intermediário (com paridade distinta), em geral de tempo de vida curto, para intermediar a dinâmica entre os estados da base computacional.}
    \label{Esquema}
\end{figure}

Existem dois principais esquemas para a escolha dos níveis internos \cite{James1998}: No primeiro, conceitualmente mais simples, os níveis internos são escolhidos tal que as funções de onda associadas aos estados tenham paridades distintas. No segundo, experimentalmente mais complexo, os níveis internos possuem a mesma paridade, mas um terceiro nível de paridade distinta deve ser usado para intermediar a interação entre os estados da base computacional. O nível intermediário, em geral um estado de tempo de vida curto, é utilizado para levar o íon rapidamente para o nível desejado inicialmente, que é escolhido para ser um estado com tempo de vida mais longo. Uma grande dessintonia entre o nível intermediário e os níveis fundamentais faz com que ele não seja efetivamente populado, resultando assim em uma interação efetiva que envolve somente os níveis fundamentais e, consequentemente, se comportando de forma equivalente a um sistema de dois níveis. Dessa forma, daqui para frente toda discussão considerará os íons como sistema de dois níveis, por simplicidade.
Ambos os esquemas estão representadas na Figura \ref{Esquema}.

 \section{Hamiltoniano de um Único Íon Aprisionado}\label{sec3}
 
 O grau de sucesso na implementação de computação quântica em sistemas físicos reais depende do controle que se pode obter sobre a dinâmica do sistema. Cada sistema físico tem métodos próprios
 a partir dos quais é possível gerar interações que permitem controlar sua dinâmica. Em plataformas de íons aprisionados, o controle dos estados que descrevem o sistema é realizado através da aplicação de pulsos de laser em frequências específicas. Para uma discussão mais profunda, esta seção tem como objetivo derivar o hamiltoniano de um único íon aprisionado em uma armadilha de Paul interagindo com um pulso de laser e, a partir deste hamiltoniano, construir as interações elementares que permitem controlar a dinâmica do sistema.
 
 O hamiltoniano a ser derivado é dado por
 \begin{equation}
    \hat{H} = \hat{H}_{m} + \hat{H}_{a} + \hat{V}_{I},
\end{equation}
\noindent onde $\hat{H}_{m}$ é o hamiltoniano do movimento do íon aprisionado na armadilha, $\hat{H}_{a}$ é o hamiltoniano do átomo ionizado e $\hat{V}_{I}$ é o potencial de interação entre o íon e o feixe aplicado. Nas próximas subseções, vamos deduzir expressões para esses hamiltonianos, quantizando diferentes aspectos do sistema.

\subsection{Quantização do movimento do íon}

Como discutido, o movimento axial do íon na armadilha é aproximadamente descrito pelo hamiltoniano do oscilador harmônico unidimensional, podendo o resultado ser generalizado para as demais dimensões. Desse modo, podemos descrever quanticamente o movimento do íon trocando-se as variáveis de coordenada e de momento pelos respectivos operadores $\hat{x}$ e $\hat{p}$ de um oscilador harmônico. Por conveniência, também definimos os operadores
\begin{subequations}
\begin{equation} \label{a}
    \hat{a}_x=\sqrt{\frac{m \nu_x}{2\hslash}}\left(\hat{x} + \frac{i}{m\nu_x}\hat{p}\right),
\end{equation}
\begin{equation} \label{ad}
    \hat{a}^{\dagger}_x=\sqrt{\frac{m \nu_x}{2\hslash}}\left(\hat{x} - \frac{i}{m\nu_x}\hat{p}\right),
\end{equation}
\end{subequations}
\noindent respectivamente chamados de operadores de aniquilação e criação do modo vibracional (axial). A fim de não sobrecarregar a notação, o subscrito da frequência secular e dos operadores de criação e aniquilação do modo serão omitidos daqui em diante, estando claro que em todo momento nos referimos aos operadores e frequência secular do modo axial. Da relação de comutação entre $\hat{x}$ e $\hat{p}$, é possível mostrar que $\comm{\hat{a}}{\hat{a}^{\dagger}}=1$. 

Invertendo as equações (\ref{a}) e (\ref{ad}), pode-se mostrar que
\begin{subequations}
\begin{equation}
   \hat{x}= \sqrt{\frac{\hslash}{2m\nu}}\left(\hat{a}^{\dagger} + \hat{a}\right),
\end{equation}
\begin{equation}
    \hat{p}=i\sqrt{\frac{\hslash m \nu}{2}}\left(\hat{a}^{\dagger} - \hat{a}\right),
\end{equation}
\end{subequations}
\noindent tal que, substituindo na versão quantizada da equação (\ref{hmclass}), obtemos
\begin{equation}\label{Hm}
    \hat{H}_m = \hslash\nu\left(\hat{a}^{\dagger}\hat{a}+\frac{1}{2}\right),
\end{equation}
\noindent isto é, o hamiltoniano do movimento do íon em termos dos operadores de criação e aniquilação, cujos significados físicos serão discutidos a seguir.

As definições de $\hat{a}$ e $\hat{a}^{\dagger}$ evidenciam a não-hermiticidade de tais operadores, mas o produto $\hat{a}^{\dagger}\hat{a}$ é hermitiano e é útil definir o operador número do modo vibracional,
\begin{equation}
    \hat{N} = \hat{a}^{\dagger}\hat{a},
\end{equation}
\noindent cuja equação de autovalores e autovetores é dada por
\begin{equation}
    \hat{N}\ket{n}=n\ket{n},
\end{equation}
\noindent onde $n$ assume apenas inteiros não-negativos \cite{Milburn1999}. Como $\comm{\hat{H}_m}{\hat{N}}=0$, $\hat{H}_m$ e $\hat{N}$ compartilham uma base comum de autoestados \cite{piza}. Desta forma, o conjunto dos autoestados do operador número forma também a base de autovetores do hamiltoniano do movimento do íon, sendo chamados de estados de Fock ou estados de número do modo.

As atuações dos operadores $\hat{a}$ e $\hat{a}^{\dagger}$ nos estados de Fock são - como esperado do tratamento do oscilador harmônico quântico - dadas por 
\begin{subequations}
\begin{equation}
    \hat{a}\ket{n} = \sqrt{n}\ket{n-1},
\end{equation}
\begin{equation}
    \hat{a}^{\dagger}\ket{n}=\sqrt{n+1}\ket{n+1},
\end{equation}
\end{subequations}
\noindent de tal forma que aplicar o operador de criação (aniquilação) no estado de número $\ket{n}$ gera um estado de número acrescido (diminuído) em uma unidade, criando (aniquilando) uma excitação no modo vibracional, i.e., um fônon. Os estados de de número, portanto, determinam a quantidade de fônons no modo.

Por fim, realocando o zero de energia na equação (\ref{Hm}), obtemos 
\begin{equation}
    \hat{H}_{m} = \hslash \nu \hat{a}^{\dagger}\hat{a}
\end{equation}
\noindent como sendo o hamiltoniano do movimento do íon em uma armadilha harmônica. O resultado pode ser generalizado para as demais dimensões, quando necessário.

\subsection{ Quantização da matéria}

Supondo que o pulso de laser aplicado é ressonante ou quase-ressonante com apenas uma transição atômica, é possível considerar os estados do íon como restritos a dois níveis de energia, desprezando a interação com todos os demais níveis, o que é de certo fundamental para o mapeamento dos estados internos na base computacional. Consideremos, dessa forma, os estados $\ket{g}$ e  $\ket{e}$, que representam os estados da base computacional  definidos anteriormente $\ket{0}$ e  $\ket{1}$ \footnote{Em textos de Mecânica Quântica, a notação $\ket{g}$ e $\ket{e}$ é amplamente utilizada. Já em textos de Computação Quântica, em um análogo da Computação Clássica, a notação $\ket{0}$ e $\ket{1}$ é a notação usual. Aqui, usaremos ambas as notações, dependendo do contexto, podendo os estados ser diretamente mapeados, isto é, $\ket{g} \longleftrightarrow \ket{0}$ e $\ket{e}  \longleftrightarrow \ket{1}$.},  com energia $E_g$ e $E_e$, respectivamente. Assim, o hamiltoniano do átomo na base de autoestados é dado por
 \begin{equation}
\hat{H}_a  = \hat{\mathbb{I}}\hat{H}_{a}\hat{\mathbb{I}} =  E_g\ketbra{g}{g} + E_e\ketbra{e}{e},
 \end{equation}
\noindent sendo $\hat{\mathbb{I}}$ o operador identidade, tal que, a partir da  representação matricial dos estados, 
\begin{align}
 \ket{g} &=
 \begin{pmatrix}
 0\\
 1\\
\end{pmatrix}, &
\ket{e} & = 
\begin{pmatrix}
 1\\
 0\\
 \end{pmatrix},
\end{align}
obtemos a forma matricial de $\hat{H}_a$,
 \begin{equation}
 \hat{H}_a = \begin{pmatrix}
    E_e & 0\\
    0 & E_g
    \end{pmatrix} ,
\end{equation}
 \noindent que pode ser escrita como 
\begin{equation}
 \hat{H}_a = \frac{1}{2}(E_g + E_e)\hat{\mathbb{I}} + \frac{1}{2}(E_e - E_g)\hat{\sigma}_{z}.
 \end{equation}
\noindent onde $\hat{\sigma}_z$ é uma das matrizes de Pauli, introduzidas no Apêndice A.

Definindo $\omega_0 = \omega_e - \omega_g$ como a frequência de transição entre o estado fundamental e o estado excitado do átomo e realocando convenientemente o zero de energia do hamiltoniano, obtemos
 \begin{equation}
 \hat{H}_a = \frac{\hslash \omega_0}{2}\hat{\sigma}_z,
 \end{equation}
\noindent como sendo o hamiltoniano do átomo ionizado. 

 \subsection{Quantização da interação entre os graus internos e externos do íon}
 
Restringindo o comprimento de onda do laser para valores muito maiores que as dimensões do íon, o acoplamento com o campo elétrico gera um dipolo elétrico induzido, cujo potencial de interação com o campo é dado por \cite{Cohen2}
\begin{equation}
   \hat{V}_I = - \hat{\bm{d}}\cdot\vb{E},
\end{equation}
\noindent onde $\hat{\bm{d}}$ é o operador de dipolo elétrico e $\vb{E}$ o campo elétrico do laser no dipolo.  
Na base de autoestados, o operador de dipolo elétrico é dado por
\begin{equation}\label{transições}
    \bm{\hat{d}} =  \sum_{i,j=g,e}  \bm{d}_{i j}\ketbra{i}{j},
\end{equation}
\noindent onde os elementos 
\begin{equation}\label{integral}
       \bm{d}_{i j} = 
       \mel{i}{\bm{\hat{d}}}{j} = - |e|  \int{ \psi_{i}^{\ast}(\vb{r})\  \vb{r}\ \psi_{j}(\vb{r}) \,\Diff3r}, 
    \end{equation}
\noindent são comumente chamados de transições de dipolo, uma vez que estão associados às transições do estado $\ket{j}$ para o $\ket{i}$, que podem ser permitidas (diferentes de zero) ou proibidas (iguais a zero) dependendo da paridade das funções de onda $\psi_i(\mathbf{r})$ e $\psi_j(\mathbf{r}$), como mostra a equação (\ref{integral}).

Assumindo que as funções de onda dos estados eletrônicos sejam funções de paridades distintas, obtemos
\begin{equation}
   \hat{V}_I = -\left(\bm{d}_{ge}\hat{\sigma}_{-} + \bm{d}_{eg}\hat{\sigma}_{+} \right)\cdot \vb{E},
\end{equation}
\noindent onde $\hat{\sigma}_{-} = \ketbra{g}{e}$  e $\hat{\sigma}_{+} = \ketbra{e}{g}$, como discutido no Apêndice A. Para campos elétricos propagantes, podemos aproximá-los por ondas planas e, portanto, podemos escrever
\begin{equation}
\hat{V}_I = -\left(\bm{d}_{ge}\hat{\sigma}_{+} + \bm{d}_{eg}\hat{\sigma}_{-}\right)\cdot \vb{E_0}\left(e^{i(\vb{k}\cdot\hat{\vb{r}} - \omega t + \phi)} +h.c.\right),
\end{equation}
\noindent onde $\vb{E_0}$ é a amplitude vetorial do campo, $\vb{k}$ é seu o vetor de onda, $\omega$ é a sua frequência de oscilação e $\phi$ é a fase da onda. $h.c.$ representa o hermitiano conjugado.

Assumindo $\bm{d}_{ge}=\bm{d}_{eg}$, que é novamente equivalente a supor que as funções de onda dos estados eletrônicos são funções reais, podemos definir a constante de acoplamento átomo-campo,
\begin{equation}
  \frac{\Omega}{2} = \abs{ \frac{e}{\hslash} \bra{g} \hat{\vb{r}}\ket{e}\cdot \vb{E_{0}}}
\end{equation}
\noindent que descreve a intensidade de acoplamento entre a transição atômica (entre os níveis eletrônicos) e o campo. Dessa forma, em termos da constante de acoplamento, podemos escrever
\begin{equation} 
\hat{V}_I = \frac{\hslash \Omega}{2}(\hat{\sigma}_{+} + \hat{\sigma}_{-}) \left(e^{i(\vb{k}\cdot\hat{\vb{r}} - \omega t + \phi)} + h.c\right)
\end{equation}
\noindent como sendo o potencial de interação átomo-campo.

\subsection{Hamiltoniano de Jaynes-Cummings}

Os resultados anteriores conduzem ao hamiltoniano total do sistema na representação de Schr\"{o}dinger,
\begin{equation}\notag
    \hat{H} =  \hslash\nu\hat{a}^\dagger\hat{a} + \frac{\hslash \omega_0}{2}\hat{\sigma}_z + \frac{\hslash \Omega}{2} (\hat{\sigma}_{+} + \hat{\sigma}_{-})\big(e^{i(\vb{k}\cdot\vb{\hat{r}} - \omega t + \phi)} + h.c.\big).
\end{equation}
\noindent O hamiltoniano de Rabi, como é conhecida a equação acima, possui soluções analíticas sofisticadas \cite{braak2011,zhong2013}. É possível, entretanto, encontrar soluções mais simples para um problema aproximado. Esta aproximação consiste em analisar a evolução temporal do hamiltoniano de Rabi e descartar os termos que oscilam mais rapidamente. A ideia é que, na média, os termos que oscilam de forma mais lenta dominam a dinâmica do sistema. Para seguir o procedimento, entretanto, é necessário reescrever o hamiltoniano de Rabi na representação de interação.

A representação de interação é uma representação intermediária entre as  representações de Schrödinger e Heisenberg \cite{piza}. Na representação de interação, tanto os estados do sistema quanto os operadores lineares que atuam sobre esses estados possuem dependência temporal. A representação de interação pode ser utilizada sempre que o hamiltoniano do sistema puder ser escrito como a soma de dois termos
\begin{equation}\
    \hat{H} = \hat{H}_{0} + \hat{H}_{1},
\end{equation}
\noindent onde, tipicamente, a separação dos termos é feita de forma que os termos mais simples do hamiltoniano total -- aqueles cujas dinâmicas são conhecidas -- sejam incluídos em $\hat{H}_{0}$ enquanto que os termos mais complicados -- cujas dinâmicas ainda desconhecemos -- sejam incluídos em $\hat{H}_1$. Em geral, os termos no último hamiltoniano descrevem a interação entre dois sistemas cujos hamiltonianos são incluídos em $\hat{H}_{0}$, conhecidos como a parte livre do sistema.

Os estados na representação de Schrödinger são conectados com os estados na representação de interação através da transformação,
\begin{equation}\label{evolução-int}
    \ket{\psi(t)}_I = \hat{U}^{\dagger}_0 (t) \ket{\psi(t)}_S = e^{i\hat{H}_{0}t/\hslash} \ket{\psi(t)}_S,
\end{equation}
\noindent enquanto a conexão entre os operadores das diferentes representações é dada por
\begin{equation}\label{operador-int}
    \hat{A}_I(t) = e^{i\hat{H}_{0}t/\hslash} \hat{A}_Se^{-i\hat{H}_{0}t/\hslash},
\end{equation}
\noindent onde os elementos da representação de Schrödinger (Interação) são representados pelo subscrito $S$ ($I$).

Relacionando os elementos de ambas as descrições, podemos derivar a equação que descreve a evolução dos estados na representação de interação. De fato, com algumas manipulações algébricas, invertendo a equação (\ref{evolução-int}) e substituindo na equação de Schrödinger, obtemos
\begin{equation}
    \
   \big({\hat{U}_0}\hat{H}\hat{U}_0 - \hat{H}_{0}\big) \ket{\psi(t)}_I = i\hslash \pdv{\ket{\psi(t)}_I}{t},
    \end{equation}
que, levando em conta a evolução dos operadores na representação de interação, equação (\ref{operador-int}), pode ser escrito como 
\begin{equation}
    \
   \hat{H}_1(t) \ket{\psi(t)}_I = i\hslash \pdv{\ket{\psi(t)}_I}{t},
    \end{equation}
\noindent que é o análogo da equação de Schr\"{o}dinger na representação de interação desde que o hamiltoniano $\hat{H} = \hat{H}_0 + \hat{H}_1$ seja representado por
\begin{equation}\label{h-interação}
     \hat{H} = \hat{U}_0^{\dagger} \hat{H}_{1}\hat{U}_0
\end{equation}
\noindent na representação de interação.

Dessa forma, a partir da equação (\ref{h-interação}), com $\hat{H}_0 = \hat{H}_a + \hat{H}_m$ e $\hat{H}_1 = \hat{V}_I$, o hamiltoniano de Rabi na representação de interação fica dado por
\begin{equation}\label{Ht}
\hat{H}  =  \frac{\hslash\Omega}{2} [\hat{\sigma}_{+}(t) + \hat{\sigma}_{-}(t)]  \left\{  e^{i\left[\eta [\hat{a}(t) + \hat{a}^{\dagger}(t)] - \omega t + \phi\right]} + h.c. \right\},
\end{equation}
\noindent onde $\eta= k\cos{\theta}\sqrt{\hslash/2 m \nu}$ é denominado o parâmetro de Lamb-Dicke, sendo $\theta$ o ângulo entre a direção do laser e a direção do movimento do íon na armadilha. A evolução dos operadores presentes no hamiltoniano de Rabi é dada por
\begin{align}
    \hat{a}^{\dagger}(t)&=\hat{a}^{\dagger}e^{i\nu t}, &  \hat{a}(t)&=\hat{a}e^{-i\nu t} ,\\
    \hat{\sigma}_{+}(t)&=\hat{\sigma}_{+}e^{i\omega_{0} t}, &   \hat{\sigma}_{-}(t)&=\hat{\sigma}_{-}e^{-i\omega_{0} t} ,
\end{align}
\noindent como é possível demonstrar pelo lema de Baker-Campbell-Haussdorff \cite{Sakurai} a partir da equação (\ref{operador-int}).

Introduzindo o operador $\hat{\gamma} = \eta\left(\hat{a}e^{-i\nu t} + \hat{a}^{\dagger}e^{i\nu t} \right)$ apenas com o intuito de simplificar a notação, podemos escrever
\begin{equation}
    \hat{H} = \frac{\hslash \Omega}{2} \left(\hat{\sigma}_{+}e^{-i\delta t}e^{i(\hat{\gamma} +\phi)} + \hat{\sigma}_{-}e^{i\Bar{\omega} t}e^{i(\hat{\gamma} +\phi)} + h.c.\right),
\end{equation}
\noindent
onde definimos $\delta \equiv \omega - \omega_0$ e $\Bar{\omega} \equiv  \omega + \omega_0$. Na aproximação de onda girante, os termos que oscilam mais rapidamente -- as exponenciais com frequência $\Bar{\omega}$ -- podem ser descartados, o que nos permite escrever
\begin{equation}
    \hat{H}_{JC} = \frac{\hslash \Omega}{2} \left(\hat{\sigma}_{+} e^{-i\delta t}  e^{i\left(\hat{\gamma} + \phi\right)}+ h.c. \right),
\end{equation}
\noindent que é conhecido como hamiltoniano de Jaynes-Cummings semiclássico.

\subsection{Interações elementares}

No desenvolvimento do hamiltoniano de Jaynes-Cummings semiclássico, o parâmetro de Lamb-Dicke, $\eta=k\cos{\theta}\sqrt{\hslash/2 m \nu}$,  foi introduzido. Este parâmetro quantifica a amplitude das oscilações do íon na armadilha quando comparado com o comprimento de onda da radiação aplicada \cite{Blatt2003}. Um regime muito útil é o regime de Lamb-Dicke, no qual $\eta \ll 1$, ou seja, a amplitude de oscilação é muito pequena se comparada com o comprimento de onda do laser aplicado. Sob estas condições, expandindo a exponencial até primeira ordem e desprezando termos de ordens superiores, temos
\begin{equation}\notag
    \hat{H}_{LD} = \frac{\hslash \Omega}{2} \left\{ \hat{\sigma}_{+} \left[  1 + i\eta \left(\hat{a}e^{-i\nu t} + \hat{a}^{\dagger}e^{i\nu t}\right)\right]e^{i(\phi - \delta t)}  + h.c.\right\},
\end{equation}
\noindent que nos permite estudar as possíveis interações que podem ser construídas alterando-se a frequência do pulso de laser aplicado, acoplando os níveis internos aos níveis externos (de vibração). A natureza dos operadores gera diferentes frequências de transição
para cada possível interação entre íon e luz que são, em geral, denominadas frequências de Rabi \cite{Blatt2004}.
    
Para $\delta = 0$, caso em que a frequência do laser aplicado é ressonante com a frequência de transição entre os níveis internos, temos
\begin{equation}
\label{singlecarrier}
\hat{H} = \frac{\hslash \Omega}{2}\left(\hat{\sigma}_{+}e^{i\phi} + \hat{\sigma}_{-}e^{-i\phi}\right),
\end{equation}
\noindent onde a aproximação de onda girante foi aplicada, nesse caso por conta de assumirmos $\nu \gg \eta \Omega$, permitindo assim desprezar os termos proporcionais às exponenciais temporais em $\hat{H}_{LD}$. A interação com o campo ressonante induz transições do tipo $\ket{g,n}\longleftrightarrow\ket{e,n}$ com frequência de Rabi $\Omega$, alterando o estado interno sem modificar o estado vibracional do modo. Essa interação é denominada interação de campo ressonante (\textit{carrier interaction}).

Para $\delta= -\nu$, o caso em que a frequência do laser aplicado é dada por $\omega= \omega_{0} - \nu$, temos
\begin{equation}
  \hat{H} = \frac{\hslash \Omega}{2}\eta\left(\hat{a}\hat{\sigma}_{+}e^{i\phi} + \hat{a}^{\dagger}\hat{\sigma}_{-}e^{-i\phi}\right),
\end{equation} 
\noindent onde a aproximação de onda girante foi aplicada (assumindo então $\nu \gg \Omega)$. A interação induz transições do tipo $\ket{g,n}\longleftrightarrow \ket{e,n-1}$ com frequência de Rabi $\Omega_{n,n-1} = \Omega\eta \sqrt{n}$. Esse hamiltoniano é conhecido como primeiro desvio para o vermelho (\textit{first red sideband interaction}) e produz interações nas quais o estado interno é excitado ao custo da aniquilação de um fônon no modo vibracional do íon, ou o estado interno é de-excitado ao custo da criação de um fônon no modo vibracional do íon. 

Para $\delta= \nu$, então $\omega= \omega_{0} + \nu$, e portanto,
\begin{equation}
 \hat{H} = \frac{\hslash \Omega}{2} \eta\left(\hat{a}^{\dagger}\hat{\sigma}_{+}e^{i\phi} + \hat{a}\hat{\sigma}_{-}e^{-i\phi}\right),
\end{equation} 
\noindent onde a aproximação de onda girante foi aplicada (novamente assumindo $\nu \gg \Omega$)
. A interação induz transições do tipo $\ket{g,n}\longleftrightarrow \ket{e,n+1}$ com frequência de Rabi $\Omega_{n,n+1} = \Omega \eta \sqrt{n+1}$. Esse hamiltoniano é conhecido como primeiro desvio para o azul (\textit{first blue sideband interaction}) e produz interações nas quais tanto o estado interno quanto o estado externo do íon sofrem excitações ou de-excitações. 

As interações construídas acima -- nomeadamente, interação de campo ressonante, primeiro desvio para o vermelho e primeiro desvio para o azul -- podem ser usadas para aplicar portas lógicas (nesse caso, rotações) sobre os estados internos de um único íon aprisionado por intermédio das matrizes $\hat{\sigma}_{+}$ e $\hat{\sigma}_{-}$, bem como de suas possíveis combinações. A engenharia de interações por trás das portas lógicas será discutida na Seção \ref{sec5}, após estudarmos as interações que podem ser geradas em uma cadeia linear de $N$ íons, isto é, um sistema de íons aprisionados que seja escalável e mais apropriado para implementação de computação quântica.

\section{Hamiltoniano de Muitos Íons Aprisionados}\label{sec4}

Processar informações e resolver problemas complexos de forma rápida e eficiente é um dos grandes desafios da computação quântica. Um importante (mas não único) aspecto acerca da capacidade de resolver problemas de um computador quântico é o número de qubits disponíveis para a implementação de algoritmos \cite{nielsen_chuang_2010}.
Em geral, problemas de interesse tecnológico e industrial demandam um alto número de qubits, principalmente quando são considerados protocolos de correção de erros \cite{Steane2006}.
Há, portanto, uma preocupação em desenvolver computadores quânticos que sejam escaláveis, isto é, que possam suportar um número progressivamente maior de qubits. Em plataformas de íons aprisionados, uma das formas de desenvolver computadores quânticos escaláveis é obter o controle da dinâmica de uma cadeia de $N$ íons, onde cada íon representa um qubit no qual se pode alocar e processar informação. Nesta seção, discutiremos a dinâmica de uma cadeia linear de $N$ íons e as possíveis interações que se pode gerar em um sistema como este, o que de certo modo representa um passo essencial para a caracterização de sistemas de íons aprisionados como arquitetura viável para implementação de computação quântica.

Para isso, consideraremos uma cadeia de íons aprisionados em uma armadilha de Paul na qual o movimento dos íons é harmônico na direção axial e fortemente limitado (a ponto de ser negligenciável) nas demais direções. Em uma configuração como esta, a repulsão coloumbiana entre os íons tem um importante papel, uma vez que o movimento de cada íon da cadeia influencia o movimento dos demais, como em um problema de osciladores acoplados. Dessa forma, descrever o movimento individual de $N$ íons acoplados em termos das coordenadas ordinárias pode se tornar uma tarefa extremamente complexa. É sempre possível, entretanto, descrever o movimento geral do sistema como uma combinação de movimentos coletivos e oscilatórios mais simples, que são denominados modos normais \cite{Goldstein}. 

Os modos normais do sistema -- um conjunto de movimentos oscilatórios coletivos cujas combinações são úteis para descrever o movimento geral do sistema -- são construídos de tal forma que sejam independentes (ou desacoplados) entre si, ainda que exista um acoplamento físico entre os íons da cadeia. Identificar cada um dos modos normais permite construir uma imagem reveladora da dinâmica do sistema, embora o movimento real seja uma combinação por vezes complicada de todos os movimentos periódicos independentes  \cite{Marion}. Por esta razão, o tratamento clássico das equações de movimento de uma cadeia linear de $N$ íons se resume a encontrar os modos normais
do nosso sistema. Posteriormente, quantizaremos os modos normais de forma semelhante ao que fizemos quando quantizamos os modos vibracionais de um único íon aprisionado na última seção. Durante a maior parte desta seção, seguiremos o tratamento matemático realizado nas referências \cite{James1998, PhysRevLett.74.4091}, considerando em vários momentos o exemplo mais simples de uma cadeia de dois íons, um problema intuitivo e até possivelmente conhecido pelo leitor.

Em um sistema como o descrito acima, é importante entender como o movimento de cada íon da cadeia -- submetidos a um potencial harmônico -- é influenciado pela repulsão coloumbiana gerada pelos demais $(N-1)$ íons. Para isso, vamos assumir que o deslocamento do $m$-ésimo íon, enumerado da esquerda para a direita, possa ser aproximado por
\begin{equation}
    x_m(t) \approx x_m^{(0)} + q_m(t),
\end{equation}
\noindent onde $x_m^{(0)}$ denota a posição de equilíbrio do $m$-ésimo íon e $q_m(t)$ denota pequenos deslocamentos em torno da posição de equilíbrio. 

Sob essa notação, a energia potencial do sistema é escrita como
\begin{equation}
      V = \sum_{m=1}^{N}\ \frac{1}{2} M \nu^2 x_m(t)^2 + \sum_{\substack{m,n=1\\ m\neq n}}^{N} \frac{Z^2 e^2}{8 \pi \epsilon_0} \frac{1}{|x_m(t) - x_n(t)|},
\end{equation}
\noindent onde $M$ é a massa de cada íon, $e$ é a carga eletrônica, $Z$ é o grau de ionização dos átomos da cadeia, $\epsilon_0$ é a permissividade elétrica do vácuo e $\nu$ é a frequência secular na direção axial. A energia cinética, a saber, é simplesmente a soma das energias cinéticas de cada um dos íons da cadeia.

Dessa forma, a posição de equilíbrio do $m$-ésimo íon da cadeia é dada por
\begin{equation}\label{equilibrio}
    \left[\pdv{V}{x_m}\right]_{x_m=x_m^{(0)}}  = 0, 
\end{equation}
\noindent e definindo o parâmetro
\begin{equation}
    \ell = \left(\frac{1}{4 \pi \epsilon_{0}}{\frac{Z^2 e^2}{ M \nu^2}}\right)^{1/3},
\end{equation}
\noindent que tem unidade de posição, podemos introduzir uma variável adimensional para a posição de equilíbrio do $m$-ésimo íon, isto é,
\begin{equation}
    u_m = x_m^{(0)}/\ell.
\end{equation}
 
 Reescrevendo a equação (\ref{equilibrio}) em termos das posições de equilíbrio adimensionais, temos um sistema linear de $N$ equações acopladas para cada $n = 1, ..., N$,
\begin{equation}\label{posição}
    u_m - \sum_{m=1}^{n-1}\frac{1}{(u_m - u_n)^2} + \sum_{m=n+1}^{N}\frac{1}{(u_m - u_n)^2} = 0,
\end{equation}
\noindent que pode ser resolvido para $N$ arbitrário. Para $N=2$, por exemplo,
\begin{align*}
x_1^{(0)} = -\frac{1}{2^{2/3}}\ell, \! && \! x_2^{(0)} =   \frac{1}{2^{2/3}} \ell.
\end{align*}

Mais do que conhecer a posição de equilíbrio, queremos descrever o movimento coletivo da cadeia de íons a partir da hamiltoniana do sistema. Para isso, a partir das energias cinética e potencial, podemos escrever a lagrangeana do sistema como
\begin{equation}
    L = \frac{M}{2}\sum_{m=1}^{N} \dot{q}_m^2 - \frac{1}{2} \sum_{m,n =1}^{N}q_m q_n\left[\pdv{V}{x_m}{x_n}\right]_{q_m=q_n=0},
\end{equation}
\noindent onde a energia potencial foi expandida em série de Taylor em torno dos pontos de equilíbrio, reajustando o zero de energia e desprezando termos de ordem superior.

Pode-se verificar que 
\begin{equation}\label{lagrangeana}
    L = \frac{M}{2}\left[\sum_{m=1}^{N} \dot{q}_m^2 - \nu^2\sum_{m,n=1}A_{mn}q_m q_n\right],
\end{equation}
\noindent onde os elementos da matriz $A_{mn}$ são dados por
\begin{equation}
 \begin{split}
A_{mn} = \begin{dcases}
   1 + 2\sum_{{\substack{p=1\\ p\neq m}}}^{N} \frac{1}{|u_m - u_p|^3}, m = n,\\ 
 \vphantom{\frac{0}{0}}  \frac{-2}{|u_m - u_n|^3}, m \neq n,
 \end{dcases}
\end{split} 
 \end{equation}
\noindent sendo simples perceber que, além de possuir elementos reais, a matriz $A$ é simétrica por consequência da ordem de derivação ser arbitrária. 

O problema agora se resume em encontrar uma transformação de coordenadas que diagonalize a matriz $A$, tornando o problema mais simples \cite{Marion}. Tais coordenadas são encontradas a partir da equação de autovalores e autovetores,
 \begin{equation}
     \sum_{n=1}^{N} A_{mn} \bm{b}_n^{(p)} = \mu_p \bm{b}_m^{(p)} \, \, \,  \, \, (p=1,...,N),
 \end{equation}
 \noindent com $\mu_p \ge 0$.  Os autovetores são enumerados pelo índice $p$  (do menor para o maior autovalor correspondente) e normalizados tal que valham as relações de completeza e ortonormalidade, 
 \begin{subequations}
  \begin{equation}\label{delta}
     \sum_{p=1}^{N} \bm{b}_m^{(p)} \bm{b}_n^{(p)}  = \delta_{mn},
 \end{equation}
   \begin{equation}
     \sum_{m=1}^{N} \bm{b}_m^{(p)} \bm{b}_m^{(q)}  = \delta_{ pq}.
 \end{equation}
 \end{subequations}

É possível demonstrar que o primeiro e o segundo autovalor e autovetor são sempre dados por
\begin{subequations}
\begin{align}
    \bm{b}^{(1)} = \frac{1}{\sqrt{N}} \begin{pmatrix} 1\\ 1 \\ \vdots \\ 1\end{pmatrix}, \quad \mu_1 = 1,
\end{align}
\begin{align}
    \bm{b}^{(2)} = \frac{1}{\sqrt{\sum_{m=1}^N u_m^2}} \begin{pmatrix} u_1\\ u_2 \\ \vdots \\ u_N\end{pmatrix}, \quad \mu_2 = 3,
\end{align}
\end{subequations}
\noindent onde as constantes $u_m$ são determinadas pela equação (\ref{posição}). Como ilustração, para $N=2$, os autovetores da matriz $A$ são 
\begin{subequations}
\begin{align*}
    \bm{b}^{(1)} = \frac{1}{\sqrt{2}} \begin{pmatrix} 1\\ 1\end{pmatrix}, \quad \mu_1 = 1,
\end{align*}
\begin{align*}
    \bm{b}^{(2)} = \frac{1}{\sqrt{2}} \begin{pmatrix} -1 \\ \phantom{0} 1\end{pmatrix}, \quad \mu_2 = 3.
\end{align*}
\end{subequations}

Os modos normais do sistema são definidos por \cite{Marion}
\begin{equation}\label{ModoNormal}
    Q_p(t) = \sum_{m=1}^{N} \bm{b}_m^{(p)} q_m(t).
\end{equation}
\noindent O primeiro desses modos, $Q_1(t)$, denominado modo do centro de massa, corresponde a um movimento no qual todos os íons oscilam para frente e para trás como se estivessem rigidamente presos uns aos outros. O segundo modo, $Q_2(t)$, denominado modo de respiração, corresponde a um movimento no qual cada íon oscila com uma amplitude proporcional à sua distância ao centro da armadilha \cite{Nagerl1998}.

Invertendo a equação (\ref{ModoNormal}), obtemos
\begin{equation}\label{q}
    q_m(t) = \sum_{p=1}^{N}\bm{b}_m^{(p)}Q_p(t),
\end{equation}
\noindent que permite reescrever a lagrangeana do sistema, equação (\ref{lagrangeana}), em termos dos modos normais,
\begin{equation}
    L = \frac{M}{2}\sum_{p=1}^{N}\left[\dot{Q}_p^2 - \nu_p^2Q_p^2\right],
\end{equation}
\noindent onde $\nu_p = \sqrt{\mu_p}\nu$ é a frequência do $p$-ésimo modo. Como discutido anteriormente, a vantagem de encontrar os modos normais que descrevem o sistema é que estes permitem estudar o movimento do sistema como uma combinação de movimentos oscilatórios independentes e de frequências bem definidas, como se vê na equação acima, da qual a hamiltoniana do sistema pode ser derivada.

 Em termos dos modos normais, a hamiltoniana da cadeia linear de $N$ íons é, considerando o momento canonicamente conjugado $P_p = M \dot{Q}_p$, dada por
\begin{equation}
    H = \sum_{p=1}^{N}\left[\frac{P_p^2}{2M} + \frac{1}{2}M\nu_p^2Q_p^{2}\right].
\end{equation}

De forma semelhante ao que foi feito no estudo da dinâmica de um único íon aprisionado, podemos introduzir os operadores de posição e momento do $p$-ésimo modo normal como combinação linear dos operadores de criação e aniquilação do respectivo modo normal, 
\begin{subequations}
\begin{equation}
   \hat{Q}_p(t)= \sqrt{\frac{\hslash}{2M\omega}}\left(\hat{a}^{\dagger}_{p} e^{i\nu_p t} + \hat{a}_{p} e^{-i\nu_p t}\right),
\end{equation}
\begin{equation}
    \hat{P}_p(t)=i\sqrt{\frac{\hslash M \omega}{2}}\left(\hat{a}^{\dagger}_{p} e^{i\nu_p t} - \hat{a}_{p}e^{-i\nu_p t}\right),
\end{equation}
\end{subequations}
\noindent já escritos na representação de interação e definidos de forma análoga àquela feita anteriormente, isto é, tal que $\comm{\hat{Q}_p}{\hat{P}_p}=i\hslash$ e $\comm{\hat{a}_{p}}{\hat{a}^{\dagger}_{p}}=1$.

A partir da equação (\ref{q}), é possível definir o operador $\hat{q}_m(t)$, que denota o deslocamento do $m$-ésimo íon em torno de sua posição de equilíbrio, em termos dos operadores de criação e aniquilação dos modos normais,
\begin{equation}\label{62}
   \hat{q}_m(t) = \sum_{p} s_m^{(p)} \left(\hat{a}^{\dagger}_{p} e^{i\nu_p t} + \hat{a}_{p} e^{-i\nu_p t}\right),
\end{equation}
\noindent onde definimos
\begin{equation}
    s_m^{(p)} = \frac{\sqrt{N}\bm{b}_m^{(p)}}{\mu_{p}^{1/4}}
\end{equation}
\noindent como sendo a constante de acoplamento do modo. Assim, para o modo do centro de massa,
\begin{equation*}
    s_m^{(1)} = 1, \, \,  \, \,  \,  \,    \,  \,  \,  \,    \,   \,  \,  \,  \,  \,  \,  \,    \nu_1 = \nu,
\end{equation*}
\noindent e para o modo de respiração,
\begin{equation}
    s_m^{(2)} = \frac{\sqrt{N}}{3^{1/4}}\frac{u_m}{(\sum_{m}^{N} u_m^2)^{\frac{1}{2}}},  \, \,  \, \, \, \,  \nu_2 = \sqrt{3}\nu.
\end{equation}

A equação (\ref{62}) expressa um importante resultado da seção. Em suma, mostramos que é possível dispor de uma estrutura organizada de íons cujas posições de equilíbrio são determinadas pela repulsão coulombiana e pelo potencial harmônico aos quais os íons estão submetidos. Nesse sistema, entretanto, a repulsão coloumbiana não permite que estudemos o movimento de cada íon separadamente, uma vez que o movimento de um íon está acoplado ao de todos os outros pela interação entre as cargas elétricas. Por outro lado, é possível descrever o deslocamento de um único íon a partir da superposição dos operadores de criação e aniquilação dos modos vibracionais coletivos. Definidos os operadores de criação e aniquilação de cada modo, o próximo passo é analisar a interação entre um único laser e a cadeia de íons aprisionados, conectando o que vimos até agora com o objetivo de aplicar portas lógicas em uma plataforma escalável de íons aprisionados cujos estados internos são mapeados na base computacional dos qubits de um computador quântico. 

\subsection{Interações elementares}

\begin{figure}[h]
    \centering
    \includegraphics[scale = 1.5]{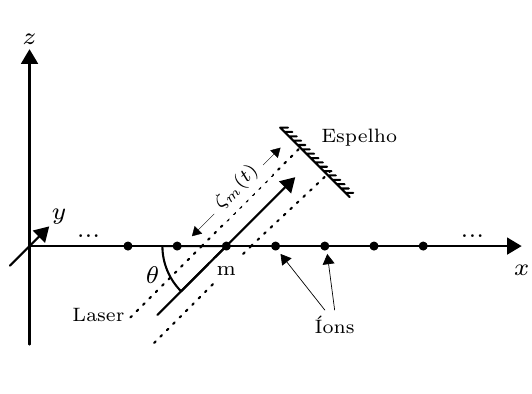}
    \caption{Representação esquemática da cadeia linear de íons proposta por Cirac e Zoller em um sistema de coordenadas cartesiano. O feixe de laser é aplicado em paralelo à direção $y$. Adaptado de \cite{James1998}.}
    \label{mirror} 
\end{figure}

Similar ao caso de um único íon aprisionado, o potencial de interação entre um pulso de laser e o $m$-ésimo íon da cadeia linear é descrito pela interação de dipolo,
\begin{equation}
   \hat{V}_I = -\bm{\hat{d}}\cdot \vb{E},
\end{equation}
\noindent onde $\hat{\bm{d}}$ é o momento de dipolo do $m$-ésimo íon e $\vb{E}$ é o campo elétrico no dipolo. Ainda seguindo a referência \cite{James1998}, considerando agora, por conveniência, um campo estacionário (onde trocamos a função exponencial que apareceu anteriormente por uma função senoidal) e realizando o mesmo procedimento de acoplamento do potencial com os estados eletrônicos discutido na seção passada, obtemos
\begin{equation}
     \hat{H} = \frac{\hslash \Omega}{2}\sin{[k\hat{\zeta}_m(t)]}e^{i(\delta t -\phi)}\hat{\sigma}_{-} + h.c.,
\end{equation} 
\noindent isto é, o hamiltoniano que descreve a interação entre o campo estacionário e o $m$-ésimo íon da cadeia na representação de interação, tendo sido aplicada a aproximação de onda girante. Como antes, $\delta$ é a dessintonia entre as frequências, $\Omega_{0}$ é a constante de acoplamento do campo com o $m$-ésimo íon, $k$ é o número de onda e $\phi$ é a fase do campo. O novo operador introduzido, $\hat{\zeta}_m(t)$, denota a distância entre o $m$-ésimo íon e o espelho plano usado para formar a onda estacionária, como esquematizado na Figura \ref{mirror}.

Ajustando a distância íon-espelho tal que a posição de equilíbrio do $m$-ésimo íon esteja em um nodo do campo,
\begin{equation}
    \hat{\zeta}_m(t) = \frac{l\lambda}{2} + \hat{q}_m(t)\cos{\theta},
\end{equation}
\noindent onde $l$ é um inteiro, $\lambda$ é o comprimento de onda do laser e $\theta$ é o ângulo entre a direção do feixe aplicado e o eixo da armadilha. Logo, a interação gerada é da forma
\begin{equation}
     \hat{H} = \frac{\hslash \Omega}{2}\sin{[k \hat{q}_m (t) \cos{\theta}]}e^{i(\delta t -\phi + l\pi)} \hat{\sigma}_{-} + h.c..
\end{equation} 

Assim, como consideramos o deslocamento do $m$-ésimo íon em torno do ponto de equilíbrio suficientemente pequeno, é possível desprezar as ordens superiores do operador $\hat{q}_m(t)$ na expansão de $\sin{[k \hat{q}_m (t) \cos{\theta}]}$, obtendo a interação
\begin{equation}\label{acopla}
\hat{H}  \approx \frac{\hslash \Omega}{2} k \cos{\theta} \hat{q}_m(t) e^{i(\delta t - \phi + l\pi)}\hat{\sigma}_{-} + h.c..
\end{equation}

Por outro lado, ajustando a distância íon-espelho tal que a posição de equilíbrio do $m$-ésimo íon esteja em um antinodo do campo,
\begin{equation}
    \hat{\zeta}_m(t) = \frac{2l - 1}{4}\lambda + \hat{q}_m(t)\cos{\theta},
\end{equation}
\noindent o hamiltoniano da interação será dado por
\begin{equation}
\hat{H} = \frac{\hslash \Omega}{2}\cos{[k\hat{q}_m(t)\cos{\theta}]}e^{i(\delta t-\phi+ l \pi/2)} \hat{\sigma}_{-} + h.c..
\end{equation}

Novamente, como consideramos o deslocamento do $m$-ésimo íon em torno do ponto de equilíbrio suficientemente pequeno, podemos aproximar $\cos{[k \hat{q}_m (t) \cos{\theta}]} \approx 1$, obtendo a interação
\begin{equation}\label{não-acopla}
\hat{H} \approx \frac{\hslash \Omega}{2}e^{i(\delta t-\phi+ l\pi/2)} \hat{\sigma}_{-} + h.c..
\end{equation}

Utilizando a equação (\ref{62}), podemos escrever o operador $\hat{q}_m(t)$ em função dos operadores de criação e destruição dos modos coletivos, tal que
\begin{equation}
    k \cos{\theta} \hat{q}_m(t) =  \frac{\eta}{\sqrt{N}} \sum_{p=1}^{N} \left(\hat{a}^{\dagger}_{p} e^{i\nu_p t} + \hat{a}_{p} e^{-i\nu_p t}\right),
\end{equation}
\noindent onde $\eta = \sqrt{\hslash k^2 cos^2{\theta}/2m\nu}$ é o parâmetro de Lamb-Dicke da interação. Dessa definição, pode-se notar que a interação construída na equação (\ref{acopla}) acopla os estados internos do $m$-ésimo íon com os estados vibracionais dos modos normais da cadeia enquanto que a interação construída na equação (\ref{não-acopla}) altera somente os estados internos do $m$-ésimo íon, sem acoplar os estados internos com os estados vibracionais dos modos coletivos.

As interações construídas nas equações (\ref{acopla}) e (\ref{não-acopla}) são essenciais para a implementação de computação quântica em sistemas de íons aprisionados, como veremos na próxima seção. Antes disso, é importante notar que a equação (\ref{acopla}), diferentemente da equação (\ref{não-acopla}), depende do número de íons na cadeia, o que pode se tornar um problema complexo uma vez que os níveis internos são acoplados com todos os $N$ modos coletivos a cada pulso aplicado no sistema. No entanto, como observado por Cirac e Zoller \cite{PhysRevLett.74.4091}, é possível obter uma condição suficiente tal que a equação (\ref{acopla}) possa ser aproximada para a interação
\begin{equation}
    \label{ciraczollercnot}
    \hat{H}_{CZ}=  \frac{\hslash \Omega \eta }{ \sqrt{4N}}\big(\hat{a}_1 e^{-i\nu_1 t} + \hat{a}_1^{\dagger} e^{i\nu_1 t}\big)e^{i(\delta t -\phi)}\hat{\sigma}_{-}+h.c.,
\end{equation}
\noindent  na qual o laser interage apenas com o modo coletivo do centro de massa, de maneira que a interação do laser com os demais modos possa ser desprezada. O hamiltoniano acima é conhecido como hamiltoniano de Cirac-Zoller. A condição na qual esta aproximação é válida é dada por
\begin{equation}
 \Bigg(\frac{\Omega\eta}{ \sqrt{N}\nu}\Bigg)^2 \ll 1
\end{equation}
\noindent e demonstrada no Apêndice B.

\section{Implementação de Portas Lógicas em Sistemas de Íons Aprisionados}\label{sec5}

Em um computador quântico, a informação é armazenada em qubits e manipulada via aplicação de portas lógicas. O objetivo dessa seção é construir, a partir das interações desenvolvidas em uma cadeia linear de $N$ íons, equações (\ref{não-acopla}) e (\ref{ciraczollercnot}), um conjunto de portas lógicas no qual quaisquer operações entre os qubits do sistema possam ser decompostas. Um conjunto como esse é chamado de conjunto universal de portas lógicas e não é único \cite{1995}. Em 1995, Sleator e Weinfurter \cite{PhysRevLett.74.4087} demonstraram que um conjunto universal de portas lógicas pode ser formado por rotações sobre um único qubit e pela porta Controlled-NOT -- uma porta quântica que atua simultaneamente em dois qubits, controlando seus respectivos estados. Em seguida, vamos construir as interações supracitadas em um sistema de $N$ íons aprisionados, construindo assim um conjunto universal de portas lógicas para a arquitetura.

\subsection{Rotações sobre um único qubit}

\begin{figure}[h]\label{bloch}
    \centering
    \includegraphics[scale = 1.2]{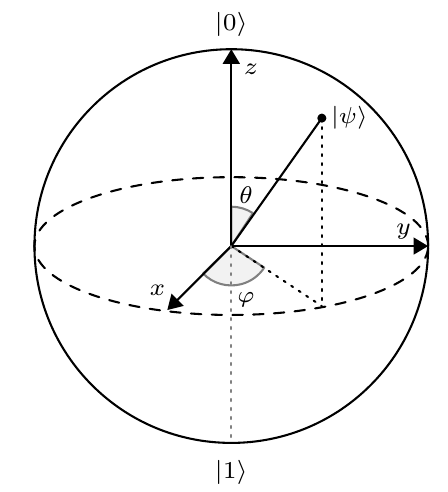}
    \caption{Representação de um estado $\ket{\psi}$ arbitrário na superfície da esfera de Bloch.}
\end{figure}

Em um sistema de íons aprisionados, o estado geral do $m$-ésimo qubit da cadeia é representado pela superposição dos níveis internos do íon,
\begin{equation}\label{qubit}
    \ket{\Psi}_m = \alpha \ket{0}_m + \beta \ket{1}_m,
\end{equation}
\noindent onde mapeamos os estados internos do íon nos estados da base computacional, sendo $\alpha$ e $\beta$ coeficientes complexos, que satisfazem $\abs{\alpha}^2 + \abs{\beta}^2 = 1$, isto é, a condição de normalização. Podemos, então, reescrever a equação (\ref{qubit}) como
\begin{equation}
    \label{qubit1}
    \ket{\Psi}_m=\cos{\theta}\ket{0}_m+e^{i\varphi}\sin{\theta}\ket{1}_m,
\end{equation}
\noindent o que nos permite representar qualquer estado de um qubit como um ponto na superfície de uma esfera de raio unitário, conhecida como esfera de Bloch \cite{nielsen_chuang_2010}, ilustrada na Figura \ref{bloch}. A esfera de Bloch oferece uma forma simples e intuitiva de visualizar o resultado de operações aplicadas em um único qubit, as quais chamamos de rotações.

A partir da representação dos estados dos qubits como pontos na superfície da esfera de Bloch, aplicar uma rotação arbitrária sobre o $m$-ésimo qubit de uma cadeia de $N$
íons é análogo a realizar uma rotação no ponto $\ket{\Psi}_{m}$, que representa o estado do qubit em relação a algum eixo arbitrário. Em outras palavras, construir uma rotação geral sobre o $m$-ésimo qubit da cadeia é construir uma interação na qual um estado da base computacional do $m$-ésimo qubit, $\ket{0}_m$ ou $\ket{1}_m$, é levado para o estado geral $\cos{\theta}\ket{0}_m+e^{i\varphi}\sin{\theta}\ket{1}_m$.

Para construir uma interação com as propriedades discutidas acima, partimos da equação (\ref{não-acopla}) com $\delta=0$, ajustando a fase do campo e aplicando a interação no $m$-ésimo qubit da cadeia por um tempo $t=k\pi/ \Omega$. Dessa forma, o operador de evolução temporal da interação se torna
\begin{equation}
    \label{Urotac}
    \hat{V}^k_m(\phi)=\exp{-ik\frac{\pi}{2}\big(\sigma_+e^{-i\phi}+\sigma_-e^{i\phi}\big)},
\end{equation}
\noindent onde $k$ é uma constante que pode ser ajustada para construir diferentes evoluções temporais. Notemos que a evolução descrita acima não atua sobre os modos coletivos, o que nos permite omitir os estados vibracionais desses modos, sabendo que não serão alterados pela interação. 

Aplicando a evolução temporal construída acima nos estados da base computacional do $m$-ésimo íon, obtemos
\begin{subequations}
\begin{equation}
\label{rotacao1}
   \ket{0}_m \stackrel{\hat{V}^k_m(\phi)}{\longrightarrow}\cos{\big(k\pi/2\big)}\ket{0}_m - ie^{i\phi}\sin{\big(k\pi/2\big)}\ket{1}_m,
\end{equation}
\begin{equation}
\label{rotacao2}
   \ket{1}_m\stackrel{\hat{V}^k_m(\phi)}{\longrightarrow}\cos{\big(k\pi/2\big)}\ket{1}_m -ie^{-i\phi}\sin{\big(k\pi/2\big)}\ket{0}_m.
\end{equation}
\end{subequations}
\noindent Assim, uma vez que os estados $\ket{0}$ e $\ket{1}$ formam a base computacional, as equações (\ref{rotacao1}) e (\ref{rotacao2}) implicam que é possível transformar um estado geral como o da equação (\ref{qubit}) em um estado como o da equação (\ref{qubit1}) a partir de aplicações de $\hat{V}_k^m(\phi)$, permitindo realizar qualquer rotação em um qubit da cadeia sem alterar os estados coletivos ou individuais dos demais qubits.

\subsection{Porta Controlled-NOT}\label{secaocnot}

A porta Controlled-NOT (ou CNOT) é uma operação condicional realizada entre dois qubits, que serão denotados pelos subscritos $m$ e $n$. Operações condicionais consistem em alterar o estado do $n$-ésimo qubit, denominado qubit alvo, dependendo do estado do $m$-ésimo qubit, denominado qubit de controle. Particularmente, a ação da CNOT pode ser descrita da seguinte forma: Se o estado do qubit de controle for $\ket{0}$ ($\ket{1}$), então o estado do qubit alvo não será (será) alterado. Em resumo:
\begin{align*}
    \ket{0}_m \ket{0}_n & \stackrel{CNOT}{\xrightarrow{\hspace*{0.8cm}}} \ket{0}_m \ket{0}_n, &  \ket{0}_m \ket{1}_n & \stackrel{CNOT}{\xrightarrow{\hspace*{0.8cm}}} \ket{0}_m \ket{1}_n ,\\
   \ket{1}_m \ket{0}_n & \stackrel{CNOT}{\xrightarrow{\hspace*{0.8cm}}}
    \ket{1}_m \ket{1}_n, &\ket{1}_m \ket{1}_n & \stackrel{CNOT}{\xrightarrow{\hspace*{0.8cm}}} \ket{1}_m \ket{0}_n.
\end{align*}
\noindent
O objetivo agora é construir um conjunto de interações que permita implementar a porta CNOT em uma cadeia linear de íons aprisionados de tal forma que, junto com as rotações aplicadas em qubits individuais, possamos formar um conjunto universal de portas lógicas no qual uma operação arbitrária entre os qubits do sistema possa ser decomposta.

\begin{figure}
    \centering
    \includegraphics[scale = 1.2]{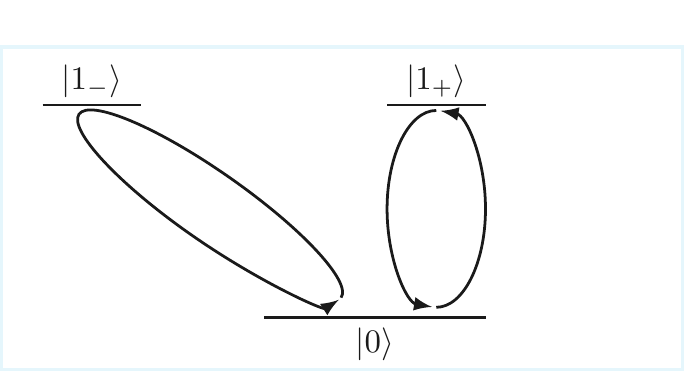}
    \caption{Representação esquemática dos níveis internos de energia de um íon aprisionado para a realização da porta \textit{CNOT}. As elipses com diferentes eixos representam diferentes polarizações de um laser com mesma frequência, utilizado para realizar transições entre os níveis. O estado $\ket{1_+}$ é o estado utilizado na base computacional e o estado $\ket{1_-}$ é utilizado como um estado auxiliar, por exemplo para se obter rotações desejadas.}
    \label{niveis_int2}
\end{figure}

Para isso, primeiramente consideremos um esquema experimental ligeiramente modificado: Cada íon possui três níveis internos, sendo dois estados excitados distintos, $\ket{1_-}$ e $\ket{1_+}$, e um estado fundamental. As transições entre o estado fundamental e os estados excitados são feitas com a mesma frequência de transição, mas com diferentes polarizações do laser incidente, como mostra a Figura \ref{niveis_int2}. Efetivamente, continuaremos lidando com um sistema de dois níveis, onde o estado $\ket{1_+}$ é mapeado no estado $\ket{1}$ da base computacional e o estado $\ket{1_-}$ é utilizado apenas como estado auxiliar para rotacionar individualmente os qubits.

Agora, consideremos o operador de evolução temporal que resulta da aplicação da interação representada pela equação (\ref{ciraczollercnot}) por um intervalo de tempo $t= k\pi \sqrt{N}/\Omega \eta$ com $\delta=-\nu_1$, isto é,
\begin{equation}
    \label{Ucnot}
    \hat{U}_m^{k,q}=\exp{-ik\frac{\pi}{2}\big(\hat{a}_1 \hat{\sigma}_{+}^q e^{-i\phi}+ \hat{a}^{\dagger}_1 \hat{\sigma}_{-}^q e^{i\phi}\big)},
\end{equation}
\noindent onde $q=+,-$ representa as duas possíveis polarizações, tendo sido a aproximação de onda girante utilizada. Aplicando a  
evolução temporal acima sobre o $m$-ésimo qubit - acompanhado do estado vibracional que será alterado pela interação, temos que
\begin{subequations}
\begin{equation*}
   \ket{0}_m\ket{0}\stackrel{\hat{U}^{k,q}_m(\phi)}{\longrightarrow}\ket{0}_m\ket{0},
\end{equation*}
\begin{equation*}
   \ket{0}_m\ket{1}\stackrel{\hat{U}^{k,q}_m(\phi)}{\longrightarrow}\cos{\big(k\pi/2\big)}\ket{0}_m\ket{1}-ie^{i\phi}\sin{\big(k\pi/2\big)}\ket{1_q}_m\ket{0},
\end{equation*}
\begin{equation*}
   \ket{1}_m\ket{0}\stackrel{\hat{U}^{k,q}_m(\phi)}{\longrightarrow}\cos{\big(k\pi/2\big)}\ket{1_q}_m\ket{0}-ie^{-i\phi}\sin{\big(k\pi/2\big)}\ket{0}_m\ket{1}.
\end{equation*}
\end{subequations}

Considerando o $m$-ésimo e $n$-ésimo íons da cadeia, a aplicação da evolução temporal 
 $\hat{U}_{CNOT}=\hat{U}_m^{1,+}\hat{U}_n^{2,-}\hat{U}_m^{1,+}$ leva a
\begin{subequations}
\label{Control-Z}
\begin{equation}
    \ket{0}_m\ket{0}_n\ket{0}\stackrel{\hat{U}_m^{1,+}\hat{U}_n^{2,-}\hat{U}_m^{1,+}(0)}{\xrightarrow{\hspace*{1.5cm}}}\ket{0}_m\ket{0}_n\ket{0},
\end{equation}
\begin{equation}
    \ket{0}_m\ket{1}_n\ket{0}\stackrel{\hat{U}_m^{1,+}\hat{U}_n^{2,-}\hat{U}_m^{1,+}(0)}{\xrightarrow{\hspace*{1.5cm}}}\ket{0}_m\ket{1}_n\ket{0},
\end{equation}
\begin{equation}
    \ket{1}_m\ket{0}_n\ket{0}\stackrel{\hat{U}_m^{1,+}\hat{U}_n^{2,-}\hat{U}_m^{1,+}(0)}{\xrightarrow{\hspace*{1.5cm}}}\ket{1}_m\ket{0}_n\ket{0},
\end{equation}
\begin{equation}
   \, \, \, \, \, \, \ket{1}_m\ket{1}_n\ket{0}\stackrel{\hat{U}_m^{1,+}\hat{U}_n^{2,-}\hat{U}_m^{1,+}(0)}{\xrightarrow{\hspace*{1.5cm}}}-\ket{1}_m\ket{1}_n\ket{0},
\end{equation}
\end{subequations}
\noindent onde ajustamos $\phi=0$ por conveniência.
À primeira vista, o estado final dessas evoluções não se parece com a porta sugerida no começo. No entanto, definindo a base de estados normalizados
\begin{equation}
    \ket{\pm}=\frac{\ket{0}\pm\ket{1}}{\sqrt{2}},
\end{equation}
\noindent o hamiltoniano especificado atua de tal maneira que
\begin{subequations}
\begin{equation}
\ket{0}_m\ket{\pm}_n\ket{0}\longrightarrow\ket{0}_m\ket{\pm}_n\ket{0},
\end{equation}
\begin{equation}
\ket{1}_m\ket{\pm}_n\ket{0}\longrightarrow\ket{1}_m\ket{\mp}_n\ket{0}.
\end{equation}
\end{subequations}
\noindent Logo, rotacionando o $n$-ésimo qubit tal que $\ket{0}\rightarrow \ket{\pm}$ e $\ket{1}\rightarrow\ket{\mp}$, é possível implementar a porta CNOT em uma cadeia de íons aprisionados. De fato, escolhendo $\phi=\pm\pi/2$ e $k=1/2$ na equação (\ref{Urotac}), 
\begin{subequations}
\begin{equation}
 \ket{0}_n\ket{0} \stackrel{\hat{V}_n^{\frac{1}{2}}(\frac{\pi}{2})}{\xrightarrow{\hspace*{1cm}}}\ket{+}_n\ket{0},
\end{equation}
\begin{equation}
  \ket{1}_n\ket{0} \stackrel{\hat{V}_n^{\frac{1}{2}}(\frac{\pi}{2})}{\xrightarrow{\hspace*{1cm}}}\ket{-}_n\ket{0},
\end{equation}
\begin{equation}
\ket{-}_n\ket{0}\stackrel{\hat{V}_n^{\frac{1}{2}}(-\frac{\pi}{2})}{\xrightarrow{\hspace*{1cm}}} \ket{1}_n\ket{0},
\end{equation}
\begin{equation}
\ket{+}_n\ket{0}\stackrel{\hat{V}_n^{\frac{1}{2}}(-\frac{\pi}{2})}{\xrightarrow{\hspace*{1cm}}} \ket{0}_n\ket{0},
\end{equation}
\end{subequations}
\noindent mostramos que o operador $\hat{V}_n^{\frac{1}{2}}\big(\frac{\pi}{2}\big)\hat{U}_{CNOT}\hat{V}^{-\frac{1}{2}}_n\big(-\frac{\pi}{2}\big)$ atua como uma porta CNOT em um estado geral. Outra forma de gerar a porta CNOT, alternativa à utilização de lasers com polarizações diferentes, podem ser encontradas em \cite{Gulde}.

\subsection{Decomposição de outras portas lógicas}

Além das representações utilizadas até o momento, as portas lógicas e estados podem também ser representados matricialmente. Na álgebra matricial, os estados $\ket{0}$ e $\ket{1}$ que formam a base computacional são definidos como
\begin{align}
 \ket{0} &\equiv 
 \begin{pmatrix}
 1\\
 0\\
\end{pmatrix}, &
\ket{1} & \equiv
\begin{pmatrix}
 0\\
 1\\
 \end{pmatrix},
\end{align}
\noindent enquanto as portas lógicas são representadas como matrizes unitárias e quadradas que atuam sobre os estados dos qubits -- superposições dos estados da base computacional, modificando-os. 

\subsubsection{Portas lógicas que atuam em um qubit}

Considerando um sistema de um único qubit, as rotações  que atuam sobre o sistema -- descritas pelas equações (\ref{rotacao1}) e (\ref{rotacao2}) -- podem ser matricialmente escritas como a matriz unitária $2\times2$
\begin{align}
\label{matrizrotac}
 R(k,\phi) &=
\begin{pmatrix}
\cos(k\pi/2) & -ie^{-i\phi}\sin(k\pi/2)\\
-ie^{i\phi}\sin(k\pi/2) & \cos(k\pi/2)\\
\end{pmatrix},
\end{align}
\noindent tal que toda operação no qubit pode ser decomposta em aplicações sequenciais da matriz de rotação. 

Dentre as principais operações aplicadas em um único qubit, é importante destacar as operações $X$, $Y$ e $Z$, que são matricialmente representadas por suas respectivas matrizes de Pauli. A fim de demonstrarmos o processo de decomposição de operações que atuam em um único qubit, vamos construir essas portas -- a menos de uma fase global $\alpha$ -- a partir da matriz de rotação $ R(k,\phi)$.

Para a execução de uma porta $X$, podemos escolher $R(1,0)$ com fase global $\alpha=\pi/2$ tal que
\begin{align}
    e^{i\pi/2}R(1,0) &=e^{i\pi/2}
    \begin{pmatrix}
    0 & -i\\
    -i & 0\\
    \end{pmatrix}
    =
    \begin{pmatrix}
    0 & 1\\
    1 & 0\\
    \end{pmatrix}.
\end{align}

 De maneira análoga, podemos gerar a  porta $Y$ utilizando $R(1,\pi/2)$ com $\alpha=\pi/2$, obtendo
\begin{align}
    Y &=e^{i\pi/2}
    \begin{pmatrix}
    0 & -1\\
    1 & 0\\
    \end{pmatrix}.
\end{align}

 Por fim, para obter a porta $Z$, podemos utilizar a relação $Z=iYX$ tal que 
\begin{align}
    Z &=e^{i\pi/2}YX=e^{i3\pi/2}
    \begin{pmatrix}
    0 & -i\\
    -i & 0\\
    \end{pmatrix}
    \begin{pmatrix}
    0 & -1\\
    1 & 0\\
    \end{pmatrix}
    =  \begin{pmatrix}
    1 & 0\\
    0 & -1\\
    \end{pmatrix}.
\end{align}

\subsubsection{Portas lógicas que atuam em dois qubits}

Para sistemas onde são manipulados emaranhamentos de $n$ qubits, a base computacional é formada por $2^n$ matrizes colunas de $2^n$ entradas e as portas lógicas são matrizes $2^n\times 2^n$ unitárias. No caso de um sistema composto por dois qubits, as matrizes que formam a base computacional são

\begin{align*}
\ket{00} &\equiv
\begin{pmatrix}
1\\
0\\
\end{pmatrix}\otimes
\begin{pmatrix}
1\\
0\\
\end{pmatrix}, &
\ket{01} &\equiv
\begin{pmatrix}
1\\
0\\
\end{pmatrix}\otimes
\begin{pmatrix}
0\\
1\\
\end{pmatrix}, &
\end{align*}
\begin{align}
\ket{10} &\equiv
\begin{pmatrix}
0\\
1\\
\end{pmatrix}\otimes
\begin{pmatrix}
1\\
0\\
\end{pmatrix}, &
\ket{11} &\equiv
\begin{pmatrix}
0\\
1\\
\end{pmatrix}\otimes
\begin{pmatrix}
0\\
1\\
\end{pmatrix}.
\end{align}

\begin{comment}
\begin{align*}
\ket{00} &\equiv
\begin{pmatrix}
1\\
0\\
0\\
0\\
\end{pmatrix}, &
\ket{01} &\equiv
\begin{pmatrix}
0\\
1\\
0\\
0\\
\end{pmatrix}, &
\end{align*}
\begin{align}
\ket{10} &\equiv
\begin{pmatrix}
0\\
0\\
1\\
0\\
\end{pmatrix}, &
\ket{11} &\equiv
\begin{pmatrix}
0\\
0\\
0\\
1\\
\end{pmatrix}.
\end{align}
\end{comment}

Além da base computacional, os estados de Bell $\ket{\beta_{xy}}
$

\begin{subequations}
    \begin{equation}
    \label{bell00}
    \ket{\beta_{00}}=\frac{1}{\sqrt{2}}(\ket{00}+\ket{11}),
    \end{equation}
    \begin{equation}
    \label{bell01}
    \ket{\beta_{01}}=\frac{1}{\sqrt{2}}(\ket{01}+\ket{10}),
    \end{equation}
    \begin{equation}
    \label{10}
    \ket{\beta_{10}}=\frac{1}{\sqrt{2}}(\ket{00}-\ket{11}),
    \end{equation}
    \begin{equation}
    \label{11}
    \ket{\beta_{11}}=\frac{1}{\sqrt{2}}(\ket{01}-\ket{10}),
    \end{equation}
\end{subequations}
\noindent formam uma outra base importante para a computação quântica, sendo a primeira porta a agir em muitos algoritmos quânticos, como os algoritmos de Shor \cite{Shor}, Deutsch-Jozsa \cite{Deutsch}, Grover \cite{Grover} e Bernstein \cite{Bernstein}. 

\begin{figure}
    \centering
    \includegraphics[scale = 1.5]{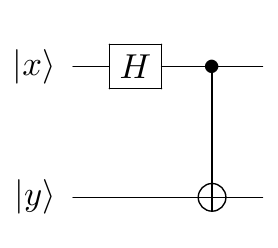}
    \caption{Circuito que representa as operações necessárias para criar o estado de Bell $\ket{\beta_{xy}}$ a partir de estados da base computacional. A primeira operação $H$ é chamada de porta de Hadamard \cite{nielsen_chuang_2010}, e pode ser decomposta como uma rotação da porta $X$. A segunda operação é uma porta CNOT, onde o qubit $\ket{x}$ está sendo controlado.}
    \label{bell}
\end{figure}

Os estados de Bell podem ser facilmente gerados a partir da base computacional utilizando a sequência de operações exibidas no circuito da Figura \ref{bell}, que envolve uma porta Hadamard, representada matricialmente por 
\begin{align}
\label{Hadamard}
    H &=
\begin{pmatrix}
1 & 1\\
1 & -1\\
\end{pmatrix} = X\,R\Big(\frac{1}{2},\frac{\pi}{2}\Big),
\end{align}
\noindent e uma porta CNOT. Dessa maneira, utilizando o método para gerar a porta CNOT obtido na seção \ref{secaocnot} e a decomposição da porta de Hadamard, é possível gerar a base formada pelos estados de Bell a partir da base computacional.

\section{Algoritmo de teletransporte quântico}\label{sec6}

Formulado por Bennett \textit{et al.} \cite{bennett1993teleporting} em 1993, o algoritmo de teletransporte quântico foi experimentalmente verificado em sistemas de fotônica \cite{Bouwmeester1997, Boschi1998} em  1997 e em sistemas de íons aprisionados por Riebe \textit{et al.} \cite{Riebe2004} e Barrett \textit{et al.} \cite{Barrett} em 2004. O protocolo desenvolvido não tem relação com o teletransporte de objetos físicos, mas sim com a transmissão de informação -- mais especificamente, a transmissão do estado de um qubit -- entre uma distância arbitrária de forma segura e eficiente. O objetivo dessa seção é discutir como o algoritmo de teletransporte quântico pode ser implementado em sistemas de íons aprisionados a partir das portas lógicas construídas nas seções anteriores, seguindo a referência \cite{Riebe2004}.

\begin{figure*}
    \centering
    \includegraphics[scale = 1.35]{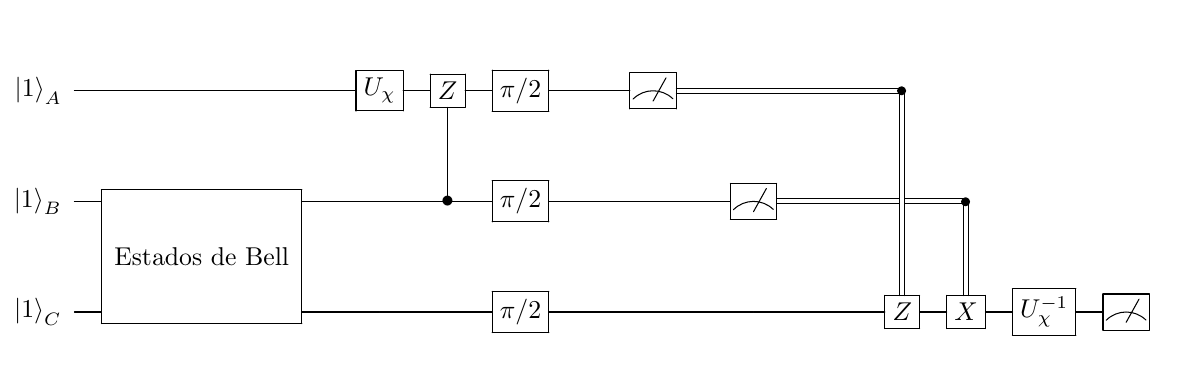}
    \caption{Algoritmo de teletransporte quântico executado por um sistema de íons aprisionados. Adaptado de \cite{Riebe2004}.}
    \label{algoritmo}
\end{figure*}

Para isso, imaginemos a seguinte situação: Alice, que possui o íon A, deseja enviar a informação contida nos estados internos do seu íon para um amigo distante, Bob, que possui o íon $C$. Alice consulta a Figura \ref{algoritmo}, que descreve o algoritmo de teletransporte quântico e percebe que precisará de um íon auxiliar, também de sua posse, o qual nomeia como íon $B$. 

Na primeira parte do algoritmo, assumimos que os íons de Alice e Bob foram previamente preparados no estado $\ket{1}_A \ket{1}_B \ket{1}_C$ e em seguida seja possível gerar o estado de superposição 
\begin{equation*}
    \ket{\Psi}=\ket{1}_A\otimes\big(\ket{0}_B\ket{1}_C+\ket{1}_B\ket{0}_C\big)/\sqrt{2},
\end{equation*}
\noindent no qual um estado de Bell  da forma explicitada na equação (\ref{bell01}) é construído entre os íons B e C. Preparada a superposição, Alice então modifica, a partir de rotações sequenciais representadas pela porta $U_\chi$, o estado do íon $A$ para o estado a ser teletransportado para Bob, $\ket{\chi}=a\ket{0}+b\ket{1}$, tal que 
\begin{equation*}
\ket{\Psi}=\big(a\ket{0}_A+b\ket{1}_A\big)\otimes\frac{(\ket{0}_B\ket{1}_C+\ket{1}_B\ket{0}_C)}{\sqrt{2}}  . 
\end{equation*}

Em seguida, uma operação controlada Hadamard-CNOT-Hadamard -- também conhecida como \textit{CZ}, uma operação do tipo $\sigma_z$ controlada pelo qubit $B$, que também pode ser realizada a partir da sequência de operações apresentadas pelas equações (\ref{Control-Z}) -- é realizada entre os íons A e B seguida pela aplicação de uma rotação $R(1/2,\pi/2)$ em cada íon. Nesse instante, o estado que descreve o sistema é dado por
\begin{eqnarray*} 
\ket{\Psi} =&& -\frac{1}{2}\ket{0}_A \ket{0}_B\left( a\ket{0}_C + b\ket{1}_C \right) \\
&&+\frac{1}{2}\ket{0}_A \ket{1}_B\left( b\ket{0}_C + a\ket{1}_C \right)\\
&&+ \frac{1}{2}\ket{1}_A \ket{0}_B\left( -a\ket{0}_C + b\ket{1}_C \right)\\
&&+ \frac{1}{2}\ket{1}_A \ket{1}_B\left( -b\ket{0}_C + a\ket{1}_C \right).
\end{eqnarray*}

O estado acima construído tem uma característica interessante que viabiliza o protocolo de teletransporte quântico. Ao realizar uma medição sobre os estados dos íons $A$ e $B$, Alice faz com que o estado do íon $C$ seja projetado em um uma superposição dos estados $\ket{0}_C$ e $\ket{1}_C$, cujos coeficientes são relacionados ao estado que se deseja teletransportar. Essa nova superposição pode então ser medida por Bob -- ante a aplicação das operações condicionais $Z$ e $X$ que finalizam o algoritmo -- tal que, com um tratamento estatístico das medições, as amplitudes do estado $\ket{\chi}$ antes conhecidas somente por Alice podem agora ser conhecidas por Bob, tendo sido a informação teletransportada com sucesso. É importante notar que, por conta das operações sequenciais realizadas no íon $A$ e a detecção realizada, o estado original é apagado do íon $A$ ao final da execução do algoritmo, o que pode ser visto como uma consequência do Teorema da não-clonagem \cite{nielsen_chuang_2010}. Como forma de confirmar o sucesso do algoritmo, é possível aplicar a porta $U_\chi^{-1}$ no estado teletransportado para Bob, esperando que se obtenha o estado $\ket{1}$ a menos de uma fase global.

\section{Estado da Arte}\label{sec7}

A arquitetura para computação quântica baseada em íons aprisionados consiste em um sistema composto de muitos átomos, que encontram-se confinados em uma determinada região do espaço por armadilhas eletromagnéticas. Embora tal plataforma apresente resultados promissores, possíveis pelo alto grau de controle sobre a cadeia de íons e a elevada precisão nos pulsos de laser aplicados para a realização de portas lógicas, sistemas de íons aprisionados encontram-se no centro de importantes pesquisas que visam melhorar as condições para sua implementação e para o desenvolvimento de computadores quânticos para uso industrial e comercial. 

\begin{figure}[t]
 \includegraphics[scale = 0.45]{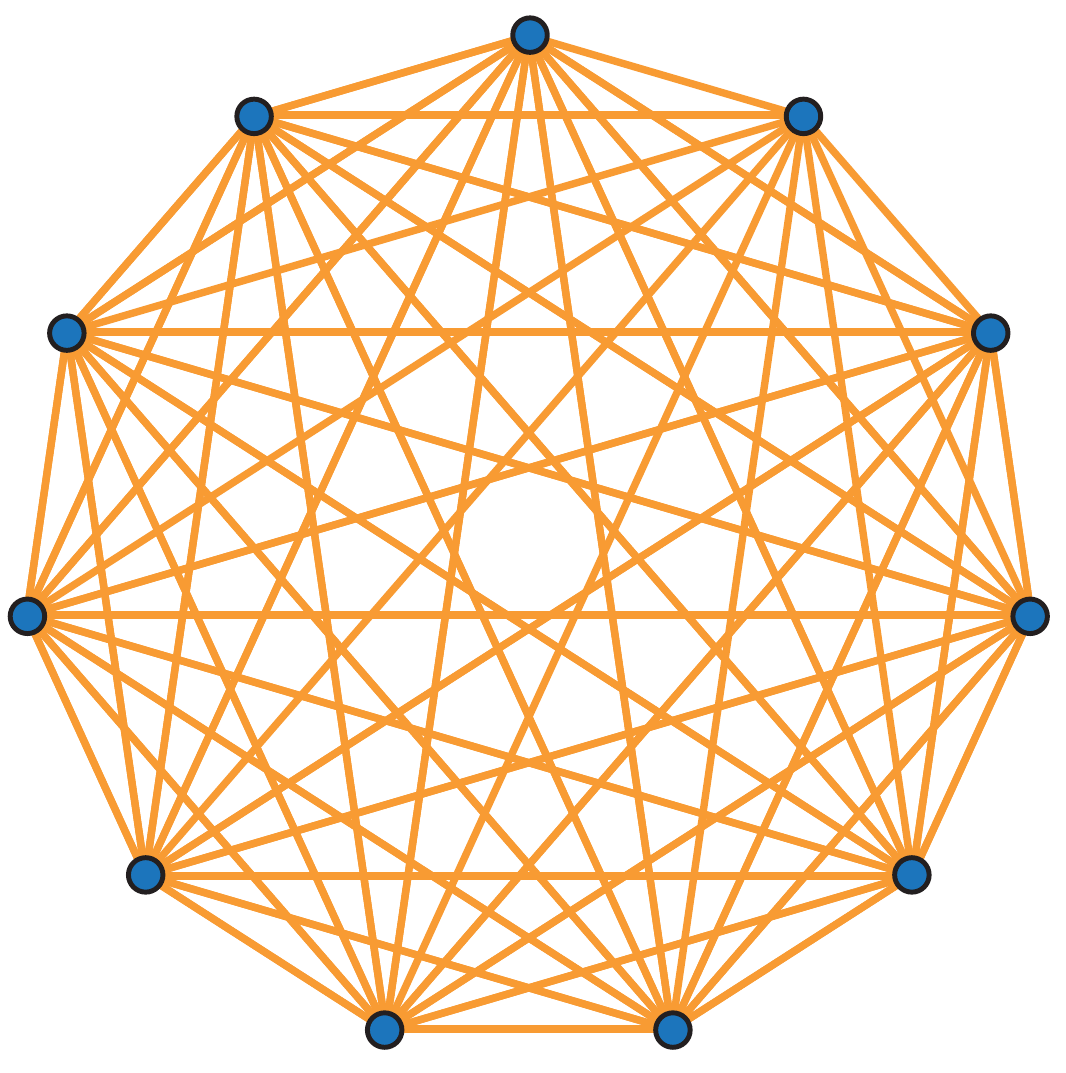}
\caption{\label{ionq-topologia} Topologia do computador quântico da empresa americana IonQ com 11 qubits totalmente interconectados. Extraído de \cite{braket}.}
\end{figure}

Grande parte dos avanços alcançados em plataformas de íons consiste em otimizar e aumentar o grau de controle de operações já implementáveis. Como exemplos atuais de otimização de operações e controle, podem ser citados métodos para reduzir os custos de portas que produzem emaranhamento entre dois \textit{qubits} com a utilização de pulsos otimizados que requerem menos energia em certos regimes de parâmetros \cite{Blumel2021}, procedimentos para emaranhar pares arbitrários de íons em uma cadeia de maneira eficiente \cite{Grzesiak2020} utilizando pulsos de laser com amplitude modulada aplicados diretamente a íons individuais da cadeia, 

Além dos avanços tecnológicos, há um grande interesse em disponibilizar arquiteturas de íons aprisionados para uso comercial. Nesse quesito, se destaca a empresa americana IonQ \cite{IonQ}, pioneira na área, que utiliza átomos de itérbio (Yb) em seus dispositivos. Fundada em 2015, a IonQ disponibiliza acesso a seus computadores quânticos por meio das plataformas Amazon Braket \cite{braket} e Azure Quantum \cite{azure} das gigantes tecnológicas Amazon e Microsoft, respectivamente. Os computadores quânticos disponibilizados nas plataformas possuem 11 qubits com topologia totalmente interconectada, como mostra a Figura \ref{ionq-topologia}, permitindo que uma porta de dois qubits atue em qualquer par de qubits da cadeia, um recurso exclusivo de plataformas de íons. A fidelidade média na aplicação das portas lógicas é acima de 99,5\% para portas atuando em um único qubit e acima de 97,5\% para portas que atuam em dois qubits. Mais informações técnicas sobre o computador quântico e implementações de algoritmos podem ser encontradas em \cite{Wright2019}, demonstrando o funcionamento (e interconectividade) da máquina. 

Recentemente, a IonQ anunciou o futuro lançamento de uma máquina com 32 qubits interconectados, previsto para o ano de 2022 \cite{32qubits}. Em seu \textit{roadmap} público \cite{roadmap}, existe a previsão de lançar um computador quântico com 256 qubits até 2026 e com 1024 qubits até 2028. Entretanto, apesar dos resultados altamente promissores mostrados pelos sistemas de íons aprisionados, ainda existem dificuldades e desafios que devem ser discutidos quando consideramos o prospecto de um computador quântico baseado em íons aprisionados de uso prático, com dezenas de milhares de qubits, incluindo protocolos de correção de erros e as decoerências do sistema. Entre esses desafios, o de maior relevância é aumentar progressivamente o número de qubits nas armadilhas sem perder o controle individual e, portanto, a alta fidelidade na aplicação de portas lógicas demonstrada pelos computadores quânticos com poucos íons.  Possíveis soluções para diferentes modelos de computadores quânticos baseados em arquiteturas de íons aprisionados, incluindo a exposta neste artigo, podem ser encontrados em (Sec. IV, \cite{Bruzewicz2019}). Uma dessas soluções, por exemplo, é compartimentar a cadeia linear em cadeia menores, divididas em módulos, que tem como consequências as dificuldades de manipular cada módulo sem prejudicar a fidelidade obtida com as portas lógicas e evitar qualquer outro tipo de perda de informação quântica.

Por fim, entre as iniciativas em desenvolvimento que atraem atenção para a área, vale ressaltar o projeto intergovernamental AQTION (\textit{Advanced Quantum computing with Trapped IONs}) \cite{AQTION} que tem como objetivo construir um computador quântico baseado em íons aprisionados com uma previsão de arrecadamento de 1 bilhão de euros entre 2018 e 2028. Em 2021, um artigo publicado pela iniciativa no periódico científico \textit{PRX Quantum} demonstrou a capacidade de construir um estado quântico com 24 íons emaranhados em um computador quântico que ocupa o espaço de apenas $1,7 \, m^3$ \cite{Pogorelov2021}, o que representa um primeiro passo para a miniaturização dos componentes necessários na engenharia de tais tecnologias.

\section{Conclusões}\label{sec8}

Neste artigo, tivemos como objetivo apresentar as características que garantem aos sistemas de íons aprisionados a condição de arquitetura para implementação de computação quântica. Para isso, considerando o modelo circuital de computação quântica, discutimos como os níveis internos de um único íon aprisionado podem ser mapeados na base computacional de um \textit{qubit} e como pulsos de laser podem ser usados para gerar interações que atuam como portas lógicas no sistema. Com a intenção de demonstrar a facilidade com que a arquitetura pode ser escalada até certo limite, expandimos a discussão para uma cadeia de $N$ íons aprisionados, na qual os modos normais da cadeia servem como base para gerar interações que permitem conectar diferentes pares de qubits. Como aplicação, discutimos a implementação do algoritmo de teletransporte quântico. Finalmente, apresentamos o atual estado da arte da arquitetura, expondo as vantagens e atuais limitações do sistema.

Em conclusão, sistemas de íons aprisionados apresentam características que reservam grandes expectativas em relação ao futuro desta arquitetura, como a alta fidelidade na geração de portas lógicas e diferentes propostas para lidar com os desafios da escalabilidade. Ainda, a interconectividade entre os qubits é um aspecto da arquitetura que, ainda que não esteja completamente esclarecido em sua importância, tem alto potencial para favorecer cada vez mais estes sistemas, como já sinalizam altos investimentos financeiros de projetos intergovernamentais como a AQTION e de empresas comerciais como a IonQ. Nesse sentido, ademais das diferentes perspectivas de futuro para a área, plataformas de íons aprisionados nos permitem explorar possibilidades únicas na implementação de algoritmos em computadores quânticos, o que torna a arquitetura uma fonte em muitos termos exclusiva de aprendizado para a evolução, desenvolvimento e implementação de computação quântica. 

\begin{acknowledgments}
Este trabalho teve o apoio do Conselho Nacional de Desenvolvimento Científico e Tecnológico (CNPq), da Coordenação de Aperfeiçoamento de Pessoal de Nível Superior (CAPES) - Código 001, e do Instituto Nacional de Ciência e Tecnologia para Informação Quântica (INCT-IQ/CNPq), Processo No. 465469/2014-0. C.J.V.-B. também agradece o apoio da Fundação de Pesquisa do Estado de São Paulo (FAPESP), Processo No. 2019/11999-5 e 2019/13143-0, e do CNPq, Processo No. 311612/2021-0.
\end{acknowledgments}

\section*{Apêndice A - Matrizes de Pauli}

As matrizes de Pauli ($\hat{\sigma}_x, \hat{\sigma}_y, \hat{\sigma}_z$) formam um conjunto de três matrizes $2 \times 2$ hermitianas que estão originalmente relacionadas com a interação entre uma partícula de spin $1/2$ e um campo eletromagnético \cite{Sakurai}, e são representadas por 
\begin{align}
\notag
\hat{\sigma}_x&= \begin{pmatrix} 0 & 1\\ 1 & 0 \end{pmatrix}, &
\hat{\sigma}_y&= \begin{pmatrix} 0 & -i\\ i & 0 \end{pmatrix},&
\hat{\sigma}_z&= \begin{pmatrix} 1 & 0\\ 0 & -1 \end{pmatrix}.
\end{align}
\noindent 

Junto da matriz identidade $2 \times 2$, as matrizes de Pauli formam a álgebra de spin $1/2$, isto é, geram o espaço dos operadores que atuam no espaço de Hilbert dos sistemas de dois níveis. Dessa forma, qualquer operador que atua nos qubits de um computador quântico pode ser mapeado em combinações desse conjunto de matrizes. Nesse sentido, vale
\begin{align}\notag
    \ketbra{g}{g} + \ketbra{e}{e} & \rightarrow \hat{\mathbb{I}}, &  \ketbra{g}{e} + \ketbra{e}{g} & \rightarrow \hat{\sigma}_x,\\
    i(\ketbra{g}{e} - \ketbra{e}{g}) & \rightarrow \hat{\sigma}_y, &   \ketbra{e}{e} - \ketbra{g}{g} & \rightarrow \hat{\sigma}_z,
\end{align}
\noindent como deve ser simples verificar. Além disso, a partir da combinação das matrizes de Pauli
\begin{equation}
\hat{\sigma}_{\pm} = \frac{\hat{\sigma}_x \pm i\hat{\sigma}_y}{2},
\end{equation}
\noindent é possível mapear
\begin{align}\notag
    \ketbra{e}{g}  & \rightarrow \hat{\sigma}_{+}, &  \ketbra{g}{e}  & \rightarrow \hat{\sigma}_{-},
\end{align}
onde $\hat{\sigma}_+$ e $\hat{\sigma}_-$ atuam como operadores de excitação e de-excitação dos qubits, como podemos ver a partir de suas respectivas definições.

\section*{Apêndice B - Hamiltoniano de Cirac-Zoller}

O objetivo deste apêndice é encontrar a condição necessária para que a interação que acopla o estado interno do $m$-ésimo íon aos $N$ estados coletivos do movimento dos íons da cadeia,
\begin{equation}
    \hat{H}  =  \frac{\hslash\Omega  \eta}{\sqrt{4 N}} \sum_{p=1}^{N} \left(\hat{a}^{\dagger}_{p} e^{i\nu_p t} + \hat{a}_{p} e^{-i\nu_p t}\right) e^{i(\delta t-\phi)}\hat{\sigma}_{-} + h.c. ,
\end{equation}
\noindent possa ser aproximada para o hamiltoniano de Cirac-Zoller,
\begin{equation}
    \label{Htmi}
  \hat{H}_{CZ}=  \frac{\hslash \Omega\eta }{\sqrt{4N}}\big(\hat{a}_1 e^{-i\nu_1 t} + \hat{a}_1^{\dagger} e^{i\nu_1 t}\big)e^{i(\delta t -\phi)}\hat{\sigma}_{-}+h.c,
\end{equation}
\noindent que acopla os estados internos do $m$-ésimo íon apenas ao modo vibracional do centro de massa da cadeia. A condição na qual essa aproximação é válida foi introduzida por James \cite{James1998} e é apresentada adiante.

Para isso, assumimos que a função de onda do \textit{m}-ésimo íon da cadeia interagindo com um laser é uma superposição do acoplamento dos estados internos com os $N$ modos vibracionais coletivos que pode ser expressa como
\begin{equation}
     \begin{split}
\ket{\Psi(t)} &=  \alpha_0(t)\ket{g}\ket{0}+\beta_0(t)\ket{e}\ket{0}+ \\
&  +\sum_{p=1}^N \alpha_p (t)\ket{g}\ket{1_p}+\sum_{p=1}^N \beta_p(t)\ket{e}\ket{1_p},
\end{split}
\end{equation}
\noindent onde escolhemos a ordem dos kets de tal forma que os kets dos estados internos, $\ket{g}$ e $\ket{e}$, são representados na frente dos estados vibracionais.

Dessa forma, a equação de movimento da função de onda que descreve o comportamento do $m$-ésimo íon é, portanto, dada por
\begin{equation}
    i\hslash \frac{\partial}{\partial t}\ket{\Psi(t)}=\hat{H}\ket{\Psi(t)},
\end{equation}
\noindent a partir de onde, escolhendo $\delta=-\nu_1$ para que o laser esteja em ressonância com o modo do centro de massa, é possível obter
\begin{equation}
\dot{\alpha}_0=\frac{\Omega\eta}{\sqrt{N}}\sum_{p=1}^{4N} s_m^{(p)}\beta_p(t),   
\end{equation}
\begin{equation}
     \dot{\beta}_0=\frac{\Omega\eta}{\sqrt{4N}}\sum_{p=1}^N s_m^{(p)}\alpha_p (t) ,
\end{equation}
\begin{equation}
    \dot{\alpha}_p(t)=-i(\nu_p-\nu_1)\alpha_p-\frac{\Omega\eta}{\sqrt{4N}}s_m^{(p)}\beta_p(t),
\end{equation}
\begin{equation}
    \dot{\beta}_p=-i(\nu_p+\nu_1)\beta_p-\frac{\Omega\nu}{\sqrt{4N}}s_m^{(p)}\alpha_0(t).
\end{equation}

Ademais, sabemos que
\begin{equation}
    \sum_{p=0}^N |\alpha_p(t)|^2+|\beta_p(t)|^2=1,
\end{equation}
\noindent o que nos permite escrever, usando a desigualdade triangular,
\begin{equation}
    |(\nu_p-\nu_1)\alpha_p(t)|=\big\vert \dot{\alpha}_p+\frac{\Omega\eta}{\sqrt{4N}}s_m^{(p)}\beta_0(t) \big\vert\leq 2\frac{\Omega \eta}{\sqrt{4N}}|s_m^{(p)}|,
\end{equation}
\begin{equation}
    |\beta_p(t)|\leq \frac{\Omega_0 \eta}{\sqrt{N}(\nu_p+\nu_1)}|s_m^{(p)}|,
\end{equation}
\noindent de maneira que a probabilidade de se encontrar o íon fora do modo do centro de massa é dada por
\begin{align}
    P_m=&\sum_{p=2}^N |\alpha_p(t)|^2+|\beta_p(t)|^2\\ &\leq 2\bigg(\frac{\Omega\eta}{\sqrt{N}\nu}\bigg)^2\sum_{p=2}^N \frac{\mu_p+1}{(\mu_p-1)^2}|s_m^{(p)}|^2.
\end{align}
No entanto, esta probabilidade está relacionada apenas ao $m$-ésimo íon da cadeia e será diferente para cada íon. Devemos considerar a probabilidade média em toda a cadeia, que é dada por
\begin{equation*}
    \overline{P}=\frac{1}{N}\sum_{m=1}^N P_i=\frac{2}{N}\Bigg(\frac{\Omega\eta}{\sqrt{N}\nu}\Bigg)^2\Bigg[\sum_{m,p}\frac{\mu_p+1}{(\mu_p-1)^2}|s_m^{(p)}|^2\Bigg]
\end{equation*}
\begin{equation*}
    \leq 2\Bigg(\frac{\Omega\eta}{\sqrt{N}\nu}\Bigg)^2\sum_{p=2}^N \frac{\mu_p+1}{(\mu_p-1)^2}\sup_{m,t}|s_m^{(p)}|^2=2\Bigg(\frac{\Omega\eta}{\sqrt{N}\nu}\Bigg)^2\Sigma(N),
\end{equation*}
\noindent onde
\begin{equation}
    \Sigma(N)=\sum_{p=2}^N \frac{\mu_p +1}{\sqrt{\mu_p}(\mu_p-1)^2}.
\end{equation}

 Numericamente, é possível mostrar que $\Sigma(N)$ é crescente e tende à $0.82$ para $N$ suficiente grande. Logo,
\begin{equation}
    \label{Pmedio}
    \overline{P}\leq 1,69\Bigg(\frac{\Omega\eta}{\sqrt{N}\nu}\Bigg)^2
\end{equation}
\noindent deve ser muito menor que a unidade para que a aproximação de Cirac e Zoller seja válida.

\end{document}